\documentclass[10pt,showpacs,preprintnumbers,amsmath,amssymb,aps,prd,nofootinbib,eqsecnum,letterpaper]{revtex4}

\usepackage{epsfig}
\usepackage{graphicx,epsf}
\usepackage{color}
\usepackage{bm}
\usepackage{psfrag}

\def\be{\begin{equation}}
\def\ee{\end{equation}}
\def\beb{\begin{equation*}}
\def\eeb{\end{equation*}}
\def\bea{\begin{eqnarray}}
\def\eea{\end{eqnarray}}
\def\beab{\begin{eqnarray*}}
\def\eeab{\end{eqnarray*}}
\def\nn{\nonumber}
\def \ni {\noindent}

\def\p{\partial}

\def\X{{{\cal{X}}}}

\def\H{{\cal H}}
\def\vp{{{\varphi}}}

\def\cs2{c_{\rm{s}}^2}
\def\dij{\delta_{ij}}

\def \beg {\begin{enumerate}}
\def \en {\end{enumerate}}
\def\fg{{\rm{flat}}}
\def\vpb{\varphi_0}
\def\Pb{P_0}
\def\rhob{\rho_0}

\def\delrho{{\delta\rho}}
\def\delP{{\delta P}}
\def\dPn{{\delta P_{\rm{nad}}}}

\def\cs{c_{\rm{s}}^2}
\newcommand\eq[1]{Eq.~(\ref{#1})}
\newcommand\eqs[1]{Eqs.~(\ref{#1})}
\newcommand{\dvp[1]}{\delta\vp_{#1}}

\newcommand{\dU[1]}{{\delta U_{#1}}}

\def\fg{{\rm{flat}}}

\def\udg{{\delta\rho}}

\def\wt{\widetilde}
\def\dT{{\delta{\bf T}_1}}
\def\dTT{{\delta{\bf T}_2}}
\def\drho{{\delta\rho_1}}
\def\drhorho{{\delta\rho_2}}
\def\drhorhorho{{\delta\rho_3}}
\def\dP{{\delta P_1}}
\def\dPP{{\delta P_2}}
\def\dPPP{{\delta P_3}}

\def\dij{\delta_{ij}}
\def\ep{\epsilon}
\newcommand\gami[1]{{\gamma_{{#1}}^{~i}}}
\newcommand\gamk[1]{{\gamma_{{#1}}^{~k}}}

\begin{document}
\title{Practical tools for third order cosmological perturbations}
\author{Adam J.~Christopherson}
\email{a.christopherson@qmul.ac.uk}
\author{Karim A.~Malik}
\email{k.malik@qmul.ac.uk}
\affiliation{
Astronomy Unit, School of Mathematical Sciences, Queen Mary University
of London, Mile End Road, London, E1 4NS, United Kingdom}
\date{\today}
\begin{abstract}
We discuss cosmological perturbation theory at third order, deriving
the gauge transformation rules for metric and matter perturbations,
and constructing third order gauge invariant quantities. We present
the Einstein tensor components, the evolution equations for a perfect
fluid, and the Klein-Gordon equation at third order, including scalar,
vector and tensor perturbations. In doing so, we also give all second
order tensor components and evolution equations in full generality.
\end{abstract}
\pacs{98.80.Jk, 98.80.Cq
\hfill
arXiv:0909.0942}

\maketitle

\section{Introduction}

Anisotropies in the cosmic microwave background and large scale
structure provide compelling evidence that the universe is not truly
homogeneous and isotropic and therefore cannot be described completely
by the Friedmann-Robertson-Walker (FRW) spacetime. Since general
relativity is highly nonlinear, it is extremely difficult to find
exact solutions which would allow for all the inhomogeneities and
anisotropies present in the real universe. Thus one usually resorts to
perturbative techniques, considering a homogeneous FRW background and
adding inhomogeneous perturbations. This is the basis of cosmological
perturbation theory, which has become a cornerstone of modern
cosmology in the last half century (see for example
Refs.~\cite{Lifshitz,Bonnor1957, Tomita1967, Bardeen80, KS, MFB,
Durrer:1993db,Bruni:1996im,Noh:2004bc,Bartolo:2004if,Nakamura,MM2008,
MW2008}).

Linear perturbation theory is now an essential tool in the
cosmologist's toolbox, and extending the theory beyond linear order is
at present a rapidly expanding area of research: 
recent years have seen second order theory gaining the attention
it duly deserves and becoming more mature. However, one does not have
to stop there: it is worthwhile and feasible to consider perturbation
theory even beyond second order, and in this paper we study aspects of
cosmological perturbation theory at third order.

Each order in perturbation theory reveals different though
complementary aspects of the underlying fully non-linear theory, and
also gives us access to information of different data sets or to
extract more information from the same data set.
First order theory allows us to model the large scale structure of the
universe and to calculate the distribution of anisotropies in the CMB. 
Second order theory then enables us to access even more information
from these data sets, by calculating higher order observables such as
the bispectrum. Third order reveals yet even more information, by 
calculating, for example, the trispectrum.

Whereas at linear order the different types of perturbations, denoted
by their transformation behaviour on three-hypersurfaces as scalar,
vector and tensor, decouple, this is no longer the case beyond first
order. At second order we find, for
example in the energy conservation equation, couplings between first
order scalar and vector perturbations through gradients, and also the
coupling of first order tensor perturbations to each other. At third
order a new coupling occurs in the energy conservation equation,
namely the coupling of scalar perturbations to tensor perturbations.
This will allow for the calculation of yet
another different observational signature, highlighting another aspect
of the underlying full theory.\\

There has already been some work on third order theory.
For example, Ref.~\cite{Hwang:2005he,Hwang:2007wq} considered third
order perturbations of pressureless irrotational fluids as ``pure''
general relativistic correction terms to second order quantities. The
calculations focused on the temporal comoving gauge, allowing the authors to
consider only second order geometric and energy-momentum
components, and neglected vector perturbations.  
Ref.~\cite{Lehners:2009ja} includes a study of third order
perturbations with application to the trispectrum in the two-field
ekpyrotic scenario in the large scale limit.
There has also been reference in the literature of the need to extend
perturbation theory beyond second order. For example, in
Ref.~\cite{Clarkson:2009hr} UV divergences in the Raychaudhuri
equation are found when considering backreaction from averaging
perturbations to second order. The authors state that these
divergences may be removed by extending perturbation theory to third,
or higher, orders.\\

In this paper, we develop the essential tools for third order
perturbation theory, such as the gauge transformation rules for
different types of perturbation, and construct gauge invariant
quantities at third order. We 
consider perfect fluids
with non-zero pressure, including all types of perturbation, namely,
scalar, vector and tensor perturbations.  In particular allowing for
vector perturbations is crucial for realistic higher order studies, as
we have recently shown that vorticity is generated at second order in
the perturbations in all models employing non-barotropic fluids
\cite{vorticity}. Hence studying irrotational fluids at higher order
will only give partial insight into the underlying physics.
We present the energy and
momentum conservation equations for such a fluid, and also give the
components of the perturbed Einstein tensor, up to third order. All
equations are given without fixing a gauge.  We also give the Klein
Gordon equation for the scalar field minimally coupled to gravity at
third order in cosmological perturbation theory.\\

The paper is organised as follows: in the next section we give our
definitions, followed by the gauge transformation rules in section
\ref{sec:gauge}.  We construct gauge invariant variables at third order
in section \ref{sec:invariant}, present the governing evolution and constraint 
equations in section
\ref{sec:equ} and conclude with a discussion in section \ref{sec:dis}
\\

In this paper, we use conformal time, $\eta$, throughout, denoting
derivatives with respect to conformal time with a prime. The scale
factor is $a$, the comoving Hubble parameter is $\H=a'/a$ 
and the physical Hubble rate is $H=a'/a^2$.
 Greek indices, $\mu,\nu,\lambda$ run from $0,\ldots,3$,
while lower case Latin indices, $i,j,k$, take the value $1,2$, or
$3$. Covariant derivatives are denoted by a semi-colon, partial
derivatives by a comma.  The order of the perturbations is usually
denoted with a subscript immediately after a perturbed quantity, when
it is split up order by order. We assume a spatially flat FRW background
throughout.

\section{Definitions}
\label{sec:def}

The components of the covariant metric tensor are
\begin{align}
g_{00} = -a^2(1+2\phi) \,, \hskip1cm
g_{0i} = a^2B_{i} \,, \hskip1cm
g_{ij} = a^2(\dij+2C_{ij}) \,,
\end{align}
where we take the three dimensional background space to be flat with
metric $\delta_{ij}$, and include scalar, vector and tensor
perturbations. Here, $\phi$ is the lapse function and $B_{i}$ is the
vector perturbation which can be split into the gradient of a scalar,
$B_{,i}$, and a divergence-free vector, $S_i$.\footnote{Note that our
  sign convention differs to that in Ref.~\cite{Hwang:2007wq}, by a
  minus sign in front of $B_i$.} $C_{ij}$ is the spatial part of the metric perturbation,
which can be decomposed as
$C_{ij}=-\psi\dij+E_{,ij}+F_{(i,j)}+\frac{1}{2}h_{ij}$. Note that in
the uniform curvature gauge, this reduces to $C_{ij} =
\frac{1}{2}h_{ij}$, where $h_{ij}$ is the tensor, or gravitational
wave, perturbation.

The contravariant metric components are obtained by imposing the
constraint $g_{\mu\nu}g^{\nu\lambda}=\delta_{\mu}{}^{\lambda} $, to
the appropriate order. To third order this gives
\begin{align}
\label{g00up}
g^{00}&=-\frac{1}{a^2}\left(1-2\phi+4\phi^2-8\phi^3-B_kB^k+4\phi B_kB^k
+2B^{i}B^{j}C_{ij}\right) \,, \\
g^{0i} &= \frac{1}{a^2}\left(B^i-2\phi B^i-2B_k C^{ki}+4\phi^2B^i+4B_kC^{ki}\phi+4C^{kj}C_j{}^iB_k-B^kB_kB^i\right)
 \,, \\
g^{ij} &= \frac{1}{a^2}\left(\delta^{ij}-2C^{ij}+4C^{ik}C_k{}^j-B^iB^j+2\phi B^iB^j-8C^{ik}C^{jl}C_{kl}+2B^iC^{kj}B_k
+2B_kB^jC^{ik}\right)\,.
\end{align}
The perturbations can be split order by order as, e.g.,
\be
\label{tensor_split}
\phi=\phi_1+\frac{1}{2}\phi_2+\frac{1}{3!}\phi_3,
\ee
but we do not explicitly perform such a split here in order to keep
expressions compact. For example, expanding the $0-0$ component fully
order by order gives
\be
g^{00}=-\frac{1}{a^2}\left(1-2\phi_1-\phi_2-\frac{1}{3}\phi_3+\phi_1^2+8\phi_1\phi_2-8\phi_1^3
-B_{1k}B_1{}^k-B_{2k}B_1{}^k+4\phi_1B_1{}^kB_{1k}+2B_1{}^iB_1{}^jC_{1ij}\right)\,,
\ee
which when compared to \eq{g00up} illustrates why we do not split
perturbations where possible in this paper.  \\

The energy momentum tensor for a perfect fluid is
\be 
\label{eq:EMfluid}
{T^\mu}_\nu=(\rho+P)u^\mu u_\nu+P{\delta^\mu}_\nu\,,
\ee
where $P$ is the pressure, $\rho$ is the energy density and 
$u^\mu$ is the fluid four velocity, defined as 
\be
u^\mu = \frac{dx^\mu}{d\tau}\,.
\ee
The fluid velocity is subject to the constraint 
\be
u^\mu u_\mu = -1\,,
\ee
and, to third order in the perturbations, has components
\begin{align}
u^i &= \frac{1}{a}v^i \,, \\
u^0 &= \frac{1}{a}\left(1-\phi+\frac{3}{2}\phi^2-\frac{5}{2}\phi^3
+\frac{1}{2}v_kv^k+v_kB^k+C_{kj}v^kv^j-2\phi v^kB_k
-\phi v^kv_k \right)\,, \\
u_i &= a\left(v_i+B_i-\phi B_i+2C_{ik}v^k+\frac{3}{2}B_i \phi^2
+\frac{1}{2}B_iv^kv_k+B_iv^kB_k\right) \,, \\
u_0 &= -a\left(1+\phi-\frac{1}{2}\phi^2+\frac{1}{2}\phi^3
+\frac{1}{2}v^kv_k+\phi v_kv^k+C_{kj}v^kv^j\right)\,.
\end{align}

\section{Gauge Transformations}
\label{sec:gauge}

We use here the active approach to deal with gauge transformations,
where the exponential map is the starting point (we refer the reader
to the recent reviews on perturbation theory
\cite{MM2008,MW2008}
for detailed discussions of gauge issues and related matters in this
section ). This allows us to immediately write down how a
tensor ${\bf T}$ transforms once the generator of the gauge
transformation, $\xi^\mu$, has been specified. The exponential map is
then
\be
\label{Ttransgeneral}
\widetilde{\bf T}=e^{\pounds_{\xi}}{\bf T}\,,
\ee
where $\pounds_{\xi}$ denotes the Lie derivative with respect
to $\xi^\lambda$.  The vector field generating the transformation,
$\xi^\lambda$, is up to third order
\be
\label{splitxi}
\xi^\mu\equiv \epsilon \xi_1^{\mu}
+\frac{1}{2}\epsilon^2\xi_2^{\mu}
+\frac{1}{6}\epsilon^2\xi_3^{\mu} +O(\epsilon^4)\,.
\ee
The exponential map can be readily expanded and, taking into account
\eq{splitxi}, gives
\be
\label{exp_expand}
\exp(\pounds_{\xi})=1+\epsilon\pounds_{\xi_1}
+\frac{1}{2}\epsilon^2\pounds_{\xi_1}^2
+\frac{1}{2}\epsilon^2\pounds_{\xi_2}
+\frac{1}{6}\epsilon^3\pounds_{\xi_3}
+\frac{1}{6}\epsilon^3\pounds_{\xi_1}^3
+\frac{1}{4}\epsilon^3\pounds_{\xi_1}\pounds_{\xi_2}
+\frac{1}{4}\epsilon^3\pounds_{\xi_2}\pounds_{\xi_1}
+\ldots
\ee
where we have kept terms up to $O(\epsilon^3)$.
Splitting the tensor ${\bf T}$ up to third order, as given in
\eq{tensor_split}, and collecting terms of like order in $\ep$ we
find that tensorial quantities transform at zeroth, first, second, and
third order, respectively, as \cite{Mukhanov96,Bruni:1996im}
\bea
\widetilde {{\bf T}_0} &=& {\bf T}_0 \,,\\
\label{Ttrans1}
\ep\widetilde \dT
&=& \ep\dT + \ep\pounds_{\xi_1} {\bf T}_0  \,,\\
\ep^2\widetilde \dTT
\label{Ttrans2}
&=& \ep^2\left(\dTT +\pounds_{\xi_2} {\bf T}_0 +\pounds^2_{\xi_1}
{\bf T}_0 + 2\pounds_{\xi_1} \dT\right)\,,\\
\label{Ttrans3}
\ep^3\widetilde {\bf \delta T_3}
&=& \ep^3\Bigg\{{\bf \delta T_3} 
+\left(\pounds_{\xi_3}+\pounds^3_{\xi_1} +\frac{3}{2}\pounds_{\xi_1}\pounds_{\xi_2}
+\frac{3}{2}\pounds_{\xi_2}\pounds_{\xi_1}
\right){\bf T}_0 
+3\left(\pounds^2_{\xi_1}+\pounds_{\xi_2}\right)\dT
+3\pounds_{\xi_1} \dTT\Bigg\}\,.
\eea

Applying the map (\ref{Ttransgeneral}) to the coordinate functions
$x^\mu$, we get the following relationship between the coordinates of a point $p$ and
a point $q$
\be
\label{defcoordtrans}
{x^\mu}( {{q}})
= e^{\xi^\lambda \frac{\p}{\p x^\lambda}\big|_{{p}}} \
x^\mu( {{p}})\,,
\ee
where we have used the fact that when acting on scalars $\pounds_{\xi}
= \xi^{\mu} (\partial/\partial x^{\mu})$ and the partial
derivatives are evaluated at $p$.
The left-hand-side and the right-hand-side of \eq{defcoordtrans} are
evaluated at different points.
Equation (\ref{defcoordtrans}) can then be expanded up to third order
as
\bea
\label{coordtrans2}
{x^\mu}(q) &=& x^\mu(p)+\epsilon\xi_1^{\mu}(p)
+\frac{1}{2}\epsilon^2\left(\xi^{\mu}_{1,\nu}(p)\xi_1^{~\nu}(p)
+ \xi_2^{\mu}(p)\right)\nonumber\\
&&+\frac{1}{6}\epsilon^3\left[\xi_3^\mu(p)+
\left(\xi^{\mu}_{1,\lambda\beta}\xi_1^\beta
+\xi^{\mu}_{1,\beta}\xi_{1,\lambda}^\beta
\right) \xi_1^{~\lambda}(p)\right]
+\frac{1}{4}\epsilon^3\left(
\xi^{\mu}_{2,\lambda}(p)\xi_1^\lambda(p)
+\xi^{\mu}_{1,\lambda}(p)\xi_2^\lambda(p)
\right)  \,,
\eea
relating the coordinates of the points $p$ and $q$.

\subsection{Four-scalars}
\label{gauge_scalar_sec}

We now turn to four-scalars, using here the energy density $\rho$ as
an example, and study its behaviour under a gauge transformation up to
third order.
It can be readily expanded as
\be
\rho=\rho_0+\delta\rho_1+\frac12\delta\rho_2+\frac{1}{3!}\delta\rho_3\,,
\ee
and we shall now detail the transformation order by order.

\subsubsection{First order}

The Lie derivative of a scalar is simply given by
\bea
\label{lie_scalar} \pounds_{\xi}\rho&=&
\xi^\lambda\rho_{,\lambda}\,.
\eea
Before we can study the transformation behaviour of the
perturbations at first order, we split the generating vector
$\xi_1^\mu$ into a scalar temporal part $\alpha_1$ and a spatial
scalar and vector part, $\beta_{1}$ and $\gami1$, according to
\be
\label{def_xi1}
\xi_1^\mu=\left(\alpha_1,\beta_{1,}^{~~i}+\gami1\right)\,,
\ee
where the vector part is divergence-free ($\p_k\gamk1=0$).

Under a first-order transformation a four scalar, here the energy
density, $\rho$, then transforms, from \eqs{Ttrans1} and
(\ref{lie_scalar}) as,
\be
\label{rhotransform1}
\widetilde\drho = \drho + \rho_0'\alpha_1 \,.
\ee
The first-order density perturbation can be made gauge-invariant by
prescribing the first order temporal gauge or time slicing,
$\alpha_1$.

\subsubsection{Second order}

At second order, as at first order, we split the generating vector
$\xi_2^\mu$ into a scalar temporal and scalar and vector spatial part,
as
\be
\label{def_xi2}
\xi_2^\mu=\left(\alpha_2,\beta_{2,}^{~~i}+\gami2\right)\,,
\ee
where the vector part is divergence-free ($\p_k\gamk2=0$).
We then find from Eqs.~(\ref{Ttrans2}) and (\ref{lie_scalar}) that a
four scalar transforms as
\bea
\label{rhotransform2}
\widetilde\drhorho = \drhorho
+\rho_0'\alpha_2&+&\alpha_1\left(
\rho_0''\alpha_1+\rho_0'{\alpha_1}'+2\drho'\right)
+\left(2\drho+\rho_0'{\alpha_1}\right)_{,k}
(\beta_{1,}^{~~k}+\gamk1)
\,.
\eea
We now need to specify the time slicing at first and second order, and
also the spatial gauge or threading at first order, in order to render
the second order density perturbation gauge-invariant.

\subsubsection{Third order}

At third order we also split the generating vector
$\xi_3^\mu$ into a scalar temporal and scalar and vector spatial part,
as
\be
\label{def_xi3}
\xi_3^\mu=\left(\alpha_2,\beta_{2,}^{~~i}+\gami3\right)\,,
\ee
where the vector part is again divergence-free ($\p_k\gamk3=0$).
We then find from Eqs.~(\ref{Ttrans3}) and (\ref{lie_scalar}) that the
energy density transforms as
\bea
\widetilde{\delta\rho_3}
=\delta \rho_3
+\left(\pounds_{\xi_3}+\pounds^3_{\xi_1} +\frac{3}{2}\pounds_{\xi_1}\pounds_{\xi_2}
+\frac{3}{2}\pounds_{\xi_2}\pounds_{\xi_1}
\right){\rho}_0 
+3\left(\pounds^2_{\xi_1}+\pounds_{\xi_2}\right)\delta\rho_1
+3\pounds_{\xi_1} \delta\rho_2\,,
\eea
which gives
\bea
\label{rhotransform3}
\widetilde{\delta\rho_3}
&=&\delta \rho_3+\rho_0'\alpha_3 
+\rho_0'''\alpha_1^3+3\rho_0''\alpha_1\alpha_{1,\lambda}\xi_1^\lambda
+\rho_0'\left(
\alpha_{1,\lambda\beta}\xi_1^\lambda
+\alpha_{1,\lambda}\xi_{1,~\beta}^\lambda\right)\xi_1^\beta\nonumber\\
&&+3\rho_0''\alpha_1\alpha_2
+\rho_0'\frac{3}{2}\left(
\alpha_{2,\lambda}\xi_1^\lambda+\alpha_{1,\lambda}\xi_2^\lambda
\right)
+3\left(\delta\rho_{1,\lambda\beta}\xi_1^\lambda
+\delta\rho_{1,\lambda}\xi_{1,~\beta}^\lambda\right)\xi_1^\beta
+3\delta\rho_{1,\lambda}\xi_2^\lambda
+3\delta\rho_{2,\lambda}\xi_1^\lambda\,.
\eea
Similar to the second order case, we need to specify the time slicings
(at all orders), and also the spatial gauge or threading at first and
second order, in order to render the third order density perturbation
gauge-invariant.

\subsection{The metric tensor}

We now give the transformation behaviour of the metric tensor. As
above, we refer to Refs.~\cite{MM2008} and \cite{MW2008} for details
on the first and second order calculations, and focus the detailed
calculation on the third order derivation.
The starting point is again the Lie derivative, which for a covariant
tensor is given by
\bea
\label{lie_tensor} \pounds_{\xi}g_{\mu\nu}
&=&g_{\mu\nu,\lambda}\xi^\lambda
+g_{\mu\lambda}\xi^\lambda_{,~\nu}+g_{\lambda\nu}\xi^\lambda_{,~\mu}\,.
\eea
%

\subsubsection{First and second order}

The metric tensor transforms at first order, from \eqs{Ttrans2} and
(\ref{lie_tensor}), as
\bea
\label{general_gmunu1}
\wt{\delta g^{(1)}_{\mu\nu}}&=&\delta g^{(1)}_{\mu\nu}
+g^{(0)}_{\mu\nu,\lambda}\xi^\lambda_1
+g^{(0)}_{\mu\lambda}\xi^\lambda_{1~,\nu}
+g^{(0)}_{\lambda\nu}\xi^\lambda_{1~,\mu}\,,
\eea
and at second order as
\bea
\label{general_gmunu2}
\wt{\delta g^{(2)}_{\mu\nu}}&=&\delta g^{(2)}_{\mu\nu}
+g^{(0)}_{\mu\nu,\lambda}\xi^\lambda_2
+g^{(0)}_{\mu\lambda}\xi^\lambda_{2~,\nu}
+g^{(0)}_{\lambda\nu}\xi^\lambda_{2~,\mu}
+2\Big[
\delta g^{(1)}_{\mu\nu,\lambda}\xi^\lambda_1
+\delta g^{(1)}_{\mu\lambda}\xi^\lambda_{1~,\nu}
+\delta g^{(1)}_{\lambda\nu}\xi^\lambda_{1~,\mu}
\Big]\nonumber \\
&&+g^{(0)}_{\mu\nu,\lambda\alpha}\xi^\lambda_1\xi^\alpha_1
+g^{(0)}_{\mu\nu,\lambda}\xi^\lambda_{1~,\alpha}\xi^\alpha_1
+2\Big[
g^{(0)}_{\mu\lambda,\alpha} \xi^\alpha_1\xi^\lambda_{1~,\nu}
+g^{(0)}_{\lambda\nu,\alpha} \xi^\alpha_1\xi^\lambda_{1~,\mu}
+g^{(0)}_{\lambda\alpha}  \xi^\lambda_{1~,\mu} \xi^\alpha_{1~,\nu}
\Big]
\nonumber \\
&&+g^{(0)}_{\mu\lambda}\left(
\xi^\lambda_{1~,\nu\alpha}\xi^\alpha_1
+\xi^\lambda_{1~,\alpha}\xi^\alpha_{1,~\nu}
\right)
+g^{(0)}_{\lambda\nu}\left(
\xi^\lambda_{1~,\mu\alpha}\xi^\alpha_1
+\xi^\lambda_{1~,\alpha}\xi^\alpha_{1,~\mu}
\right)\,.
\eea

Using the method detailed in Refs.~\cite{MM2008} and \cite{MW2008} we
can extract from the general expressions \eqs{general_gmunu1} and
(\ref{general_gmunu2}) the transformations behaviour of a particular
metric function. We focus here on the curvature perturbation $\psi$,
and find that it transforms at first order as \cite{Bardeen80,KS}
\bea
\label{transpsi1}
\widetilde \psi_1 &=& \psi_1-\H\alpha_1 \,.
\eea
At second order we get after some calculation
\footnote{The general expression given above in \eq{general_gmunu2}
  gives the transformation for $C_{2ij}$, namely
\bea
\label{Cij2trans}
2\widetilde C_{2ij}&=&2C_{2ij}+2\H\alpha_2 \delta_{ij}
+\xi_{2i,j}+\xi_{2j,i}+\X_{ij}\,,
\eea
and hence intermediate steps are necessary to extract the
transformation of a particular component (see Refs.~\cite{MM2008} and
\cite{MW2008} for details).}
\be
\label{transpsi2}
\wt\psi_2=\psi_2-\H\alpha_2-\frac{1}{4}\X^k_{2~k}
+\frac{1}{4}\nabla^{-2} \X^{ij}_{2~,ij}\,,
\ee
where $\X_{2ij}$ contains terms quadratic in the first order
perturbations. For ease of presentation, on taking only scalar
perturbations into account and working on large scales, where we can
neglect gradients, $\X_{2ij}$ takes the simple form
\bea
\label{X2ijdef}
\X_{2ij}&\equiv&
2\Big[\left(\H^2+\frac{a''}{a}\right)\alpha_1^2
+\H\alpha_1\alpha_1'\Big] \delta_{ij}
+4\alpha_1\left(C_{1ij}'+2\H C_{1ij}\right)\,,
\eea
and \eq{transpsi2} reduces to
\be
\label{transpsi2ls}
\wt\psi_2=\psi_2-\H\alpha_2-\left(\H^2+\frac{a''}{a}\right)\alpha_1^2
-\H\alpha_1\alpha_1'+2\alpha_1\left(\psi_1'-\H\psi_1\right)\,.
\ee

\subsubsection{Third order}

As above in the case of the transformation behaviour of a four-scalar
at third order, the change under a gauge transformation of a
two-tensor can be found applying the same methods as at second
order. We therefore find that the metric tensor transforms at third
order, from \eqs{Ttrans3} and (\ref{lie_tensor}), as
\bea
\label{general_gmunu3}
\wt{\delta g^{(3)}_{\mu\nu}}&=&\delta g^{(3)}_{\mu\nu}
+g^{(0)}_{\mu\nu,\lambda}\xi^\lambda_3
+g^{(0)}_{\mu\lambda}\xi^\lambda_{3~,\nu}
+g^{(0)}_{\lambda\nu}\xi^\lambda_{3~,\mu}\nn\\
&&+3\Bigg[
\delta g^{(1)}_{\mu\nu,\lambda}\xi^\lambda_2
+\delta g^{(1)}_{\mu\lambda}\xi^\lambda_{2~,\nu}
+\delta g^{(1)}_{\lambda\nu}\xi^\lambda_{2~,\mu}
+\delta g^{(2)}_{\mu\nu,\lambda}\xi^\lambda_1 
 +\delta g^{(2)}_{\mu\lambda}\xi^\lambda_{1~,\nu}
+\delta g^{(2)}_{\lambda\nu}\xi^\lambda_{1~,\mu}\nn\\
&&+\delta g^{(1)}_{\mu\nu,\lambda\alpha}\xi^\lambda_1\xi^\alpha_1
+\delta g^{(1)}_{\mu\nu,\lambda}\xi^\lambda_{1~,\alpha}\xi^\alpha_1
 +2\Big[
\delta g^{(1)}_{\mu\lambda,\alpha} \xi^\alpha_1\xi^\lambda_{1~,\nu}
+\delta g^{(1)}_{\lambda\nu,\alpha} \xi^\alpha_1\xi^\lambda_{1~,\mu}
+\delta g^{(1)}_{\lambda\alpha}  \xi^\lambda_{1~,\mu} \xi^\alpha_{1~,\nu}
\Big]
\nonumber \\
&&+\delta g^{(1)}_{\mu\lambda}\left(
\xi^\lambda_{1~,\nu\alpha}\xi^\alpha_1
+\xi^\lambda_{1~,\alpha}\xi^\alpha_{1,~\nu}
\right)
+\delta g^{(1)}_{\lambda\nu}\left(
\xi^\lambda_{1~,\mu\alpha}\xi^\alpha_1
+\xi^\lambda_{1~,\alpha}\xi^\alpha_{1,~\mu}
\right)\Bigg]\nonumber \\
&&+\frac{3}{2}\Bigg[
2g^{(0)}_{\mu\nu,\lambda\beta}\xi_2^\lambda\xi_1^\beta
+g^{(0)}_{\mu\nu,\lambda}\left(
\xi_{2,\beta}^\lambda\xi_1^\beta+\xi_{1,\beta}^\lambda\xi_2^\beta
\right)
+2g^{(0)}_{\lambda\beta}\left(
\xi_{2,\mu}^\lambda\xi_{1,\nu}^\beta+\xi_{2,\nu}^\lambda\xi_{1,\mu}^\beta
\right)
\nonumber \\
&&+\left(
g^{(0)}_{\mu\lambda,\beta}\xi_{2,\nu}^\lambda
+g^{(0)}_{\lambda\nu,\beta}\xi_{2,\mu}^\lambda
+g^{(0)}_{\mu\lambda}\xi_{2,\beta\nu}^\lambda
+g^{(0)}_{\lambda\nu}\xi_{2,\beta\mu}^\lambda
\right)\xi_1^\beta
+\left(g^{(0)}_{\beta\nu,\lambda}\xi_2^\lambda
+g^{(0)}_{\lambda\nu}\xi_{2,\beta}^\lambda
\right)\xi_{1,\nu}^\beta \nn\\
&&+\left(g^{(0)}_{\beta\nu,\lambda}\xi_2^\lambda
+g^{(0)}_{\lambda\nu}\xi_{2,\beta}^\lambda
\right)\xi_{1,\mu}^\beta
%
+\left(
g^{(0)}_{\mu\lambda,\beta}\xi_{1,\nu}^\lambda
+g^{(0)}_{\lambda\nu,\beta}\xi_{1,\mu}^\lambda
+g^{(0)}_{\mu\lambda}\xi_{1,\beta\nu}^\lambda
+g^{(0)}_{\lambda\nu}\xi_{1,\beta\mu}^\lambda
\right)\xi_2^\beta\nn\\
&&+\left(g^{(0)}_{\beta\nu,\lambda}\xi_1^\lambda
+g^{(0)}_{\lambda\nu}\xi_{1,\beta}^\lambda
\right)\xi_{2,\nu}^\beta
+\left(g^{(0)}_{\beta\nu,\lambda}\xi_1^\lambda
+g^{(0)}_{\lambda\nu}\xi_{1,\beta}^\lambda
\right)\xi_{2,\mu}^\beta
\Bigg]\nonumber \\
&&+\Bigg\{
g^{(0)}_{\mu\nu,\alpha\beta\lambda}\xi_{1}^\alpha\xi_{1}^\beta\xi_{1}^\lambda
+3g^{(0)}_{\mu\nu,\lambda\beta}\xi_{1}^\lambda\xi_1^\alpha\xi_{1,\alpha}^\beta
+g^{(0)}_{\mu\nu,\lambda}\xi_{1,\alpha\beta}^\lambda\xi_{1}^\alpha\xi_{1}^\beta
+g^{(0)}_{\mu\nu,\lambda}\xi_{1,\beta}^\lambda\xi_{1,\alpha}^\beta\xi_{1}^\alpha
\nonumber \\
&&
+3g^{(0)}_{\mu\beta,\lambda}
\left(\xi_{1}^\lambda\xi_{1,\alpha}^\beta\xi_{1,\nu}^\alpha
+\xi_{1}^\alpha\xi_{1}^\lambda\xi_{1,\alpha\nu}^\beta\right)
+3g^{(0)}_{\mu\alpha,\lambda}
\xi_{1}^\beta\xi_{1,\beta}^\lambda\xi_{1,\nu}^\alpha
+3g^{(0)}_{\mu\alpha,\beta\lambda}
\xi_{1}^\beta\xi_{1}^\lambda\xi_{1,\nu}^\alpha
\nonumber \\
&&+g^{(0)}_{\mu\lambda}\Bigg[
\xi_{1}^\alpha\xi_{1}^\beta\xi_{1,\alpha\beta\nu}^\lambda
+\xi_{1,\alpha}^\beta\xi_{1,\beta}^\lambda\xi_{1,\nu}^\alpha
+\xi_{1}^\alpha\xi_{1,\alpha}^\beta\xi_{1,\beta\nu}^\lambda
+2\xi_{1}^\beta\xi_{1,\alpha\beta}^\lambda\xi_{1,\nu}^\alpha
+\xi_{1}^\beta\xi_{1,\alpha}^\lambda\xi_{1,\beta\nu}^\alpha\Bigg]
\nonumber \\
&&+3g^{(0)}_{\nu\beta,\lambda}
\left(\xi_{1}^\lambda\xi_{1,\alpha}^\beta\xi_{1,\mu}^\alpha
+\xi_{1}^\alpha\xi_{1}^\lambda\xi_{1,\alpha\mu}^\beta\right)
+3g^{(0)}_{\nu\alpha,\lambda}
\xi_{1}^\beta\xi_{1,\beta}^\lambda\xi_{1,\mu}^\alpha
+3g^{(0)}_{\nu\alpha,\beta\lambda}
\xi_{1}^\beta\xi_{1}^\lambda\xi_{1,\mu}^\alpha
\nonumber \\
&&+g^{(0)}_{\nu\lambda}\Bigg[
\xi_{1}^\alpha\xi_{1}^\beta\xi_{1,\alpha\beta\mu}^\lambda
+\xi_{1,\alpha}^\beta\xi_{1,\beta}^\lambda\xi_{1,\mu}^\alpha
+\xi_{1}^\alpha\xi_{1,\alpha}^\beta\xi_{1,\beta\mu}^\lambda
+2\xi_{1}^\beta\xi_{1,\alpha\beta}^\lambda\xi_{1,\mu}^\alpha
+\xi_{1}^\beta\xi_{1,\alpha}^\lambda\xi_{1,\beta\mu}^\alpha\Bigg]
\nonumber \\
&&+6g^{(0)}_{\alpha\beta,\lambda}\xi_{1}^\lambda
\xi_{1,\nu}^\beta
+3g^{(0)}_{\alpha\lambda}\Bigg[
\xi_{1,\beta}^\lambda\left(
\xi_{1,\mu}^\alpha\xi_{1,\nu}^\beta+\xi_{1,\nu}^\alpha\xi_{1,\mu}^\beta
\right)
+\xi_{1}^\beta\left(
\xi_{1,\mu}^\alpha\xi_{1,\beta\nu}^\lambda
+\xi_{1,\nu}^\alpha\xi_{1,\beta\mu}^\lambda
\right)\Bigg]
\Bigg\}
\,.
\eea
However, in this case it becomes even more obvious than in section
\ref{sec:def} that the expressions at third order are of not inconsiderable
size. This will also be clear from the Einstein tensor components and
the evolution equations given below in Section \ref{sec:equ}.

Now, following along the same lines as at second order,
\eq{general_gmunu3} gives the transformation for the spatial part of
the metric at third order,
\bea
\label{Cij3trans}
2\widetilde C_{3ij}&=&2C_{3ij}+2\H\alpha_3 \delta_{ij}
+2\xi_{3(i,j)}
+\X_{3ij}\,,
\eea
where $\X_{3ij}$ contains terms cubic in the first order
perturbations. Extracting the curvature perturbation gives
\be
\label{transpsi3}
\wt\psi_3=\psi_3-\H\alpha_3-\frac{1}{4}\X^k_{3~k}
+\frac{1}{4}\nabla^{-2} \X^{ij}_{3~,ij}\,.
\ee
This expression is general, including scalar,
vector, and tensor perturbations and is valid on all scales.
However, as above we shall detail here only the expression valid for
scalar perturbations and large scales, at third order an even bigger
blessing than at second order, and find that $\X_{3ij}$ takes then the 
simple form
\bea
\label{X3ijdef}
\X_{3ij}&\equiv&
2a^2\delta_{ij}\Bigg\{
-3\Big[\alpha_2\psi_1'+\frac{1}{2}\alpha_1\psi_2'
+\alpha_1\alpha_1'\left(\psi_1'+2\H\psi_1\right)
+\alpha_1^2\left(\psi_1''+4\H\psi_1'\right)
+2\H\alpha_2\psi_1+\H\alpha_1\psi_2
+2\left(\frac{a''}{a}+\H^2\right)\alpha_1^2\psi_1
\Big]\nonumber\\
&+&\left(\frac{a'''}{a}+3\H\frac{a''}{a}\right)\alpha_1^3
+3\left(\frac{a''}{a}+\H^2\right)\alpha_1^2\alpha_1'
+\H\alpha_1\left(\alpha_1''\alpha_1+{\alpha_1'}^2\right)
+\frac{3}{2}\left[\H\left(\alpha_1\alpha_2'+\alpha_1'\alpha_2\right)
+2\left(\frac{a''}{a}+\H^2\right)\alpha_1\alpha_2
\right]\Bigg\}\,.\nonumber \\
\eea
Hence we finally get for the transformation of $\psi_3$ 
\bea
\label{transpsi3ls}
-\wt{\psi_3}&=&-\psi_3+\H\alpha_3
+\left(\frac{a'''}{a}+3\H\frac{a''}{a}\right)\alpha_1^3
+3\left(\frac{a''}{a}+\H^2\right)\alpha_1^2\alpha_1'\nonumber \\
&&+\H\alpha_1\left(\alpha_1''\alpha_1+{\alpha_1'}^2\right)
+\frac{3}{2}\left[\H\left(\alpha_1\alpha_2'+\alpha_1'\alpha_2\right)
+2\left(\frac{a''}{a}+\H^2\right)\alpha_1\alpha_2
\right] \\
&&-3\Big[\alpha_2\psi_1'+\frac{1}{2}\alpha_1\psi_2'
+\alpha_1\alpha_1'\left(\psi_1'+2\H\psi_1\right)
+\alpha_1^2\left(\psi_1''+4\H\psi_1'\right)
+2\H\alpha_2\psi_1+\H\alpha_1\psi_2
+2\left(\frac{a''}{a}+\H^2\right)\alpha_1^2\psi_1
\Big]
\,.\nonumber
\eea
%

\section{Gauge-invariant quantities}
\label{sec:invariant}

In the previous section we have described how perturbations transform
under a gauge shift. We can now use these results to construct
gauge-invariant quantities, in particular the curvature perturbation
on uniform density hypersurfaces, $\zeta$. In this section, as before,
we consider only scalar perturbations, and restrict ourselves to the
large scale limit.\\

\subsection{Defining hypersurfaces}
\label{sec:alphas}

From
\eqs{rhotransform1}, (\ref{rhotransform2}), and (\ref{rhotransform3})
we find the time slicing defining uniform density hypersurfaces
at first, second and third order in the large scale limit as
\bea
\alpha_{1\udg}&=&-\frac{\delta\rho_1}{\rho_0'}\,,\\
\alpha_{2\udg}&=&-\frac{\delta\rho_2}{\rho_0'}
+\frac{\delta\rho_1\delta\rho_1'}{{\rho_0'}^2}\,,\\
\alpha_{3\udg}&=&-\frac{\delta\rho_3}{\rho_0'}
+\frac{1}{2{\rho_0'}^2}\Big[
3\left(\delta\rho_1\delta\rho_2'+\delta\rho_1'\delta\rho_2\right)
-\frac{\delta\rho_1''{\delta\rho_1}^2}{\rho_0'}
-4{\delta\rho_1'}^2\frac{\delta\rho_1}{\rho_0'}
+\rho_0''\delta\rho_1'\left(\frac{\delta\rho_1}{\rho_0'}\right)^2
\Big]\,.
\eea
Similarly, the temporal gauge transformation on uniform curvature hypersurfaces is
 defined by evaluating
\eqs{transpsi1}, (\ref{transpsi2ls}), and (\ref{transpsi3ls}) and gives, at first, second and third
order,
\bea
\alpha_{1\fg}&=&\frac{\psi_1}{\H}\,,\\
\alpha_{2\fg}&=&
\frac{\psi_2}{\H}-\frac{4\psi_1^2}{\H}
+\frac{\psi_1\psi_1'}{\H^2}
\,,\\
\alpha_{3\fg}&=&\frac{\psi_3}{\H}+\frac{1}{2\H^2}\Big[3\psi_1'\psi_2
+\frac{\psi_1^2\psi_1''}{\H}+6\H\psi_1\psi_2
+\frac{4\psi_1\psi_1^2}{\H}\Big]-\frac{\psi_1^2\psi_1'}{\H^4}\Big(\frac{a''}{a}-\frac{37}{2}\H^2\Big)+\frac{8\psi_1^3}{\H}\,.
\eea
%

\subsection{Constructing gauge-invariant quantities}

We can now combine the results found so far to get gauge-invariant
quantities, and as before choose the density perturbation on uniform
curvature hypersurfaces and the curvature perturbation on uniform
density slices as examples.

\subsubsection{Energy density perturbation on uniform curvature hypersurfaces}

A gauge invariant matter quantity of interest is the perturbation to
the energy density on uniform curvature hypersurfaces. This is
obtained by substituting the temporal gauge transformation components
in the uniform density gauge (given in section \ref{sec:alphas}) into
the appropriate transformation equation (that is,
Eqs.~(\ref{rhotransform1}), (\ref{rhotransform2}) or
(\ref{rhotransform3})). This gives, at first, second and third order,
respectively,
\be
\widetilde{\delta\rho_{1\fg}}=\drho+\rhob'\frac{\psi_1}{\H}\,,
\ee
\be
\widetilde{\delta\rho_{2\fg}}
=\drhorho+\rhob'\frac{\psi_2}{\H}+\frac{\psi_1^2}{\H}\left(
\frac{\rhob''}{\H}-4\rhob'-\frac{\rhob'}{\H^2}\Big(\frac{a''}{a^2}-\H^2\Big)\right)
+2\frac{\psi_1}{\H}\Big(\rhob'\frac{\psi_1'}{\H}+\drho'\Big)\,,
\ee
\begin{align}
\widetilde{\delta\rho_{3\fg}} &=
\drhorhorho+\rhob'\frac{\psi_3}{\H}+\frac{3\rhob'}{2\H^2}(2\psi_2\psi_1'+\psi_2'\psi_1)
+3\frac{\psi_1^2}{\H^3}\Big[2(\rhob'\psi_1+\psi_1'\rhob'')+\psi_1''\Big]
+3\frac{\psi_2\psi_1}{\H^2}\Big[\rhob''+2\rhob'\H-\rhob'\frac{a''}{a}\Big]\nn\\
&
-9\rhob'\frac{\psi_1^2\psi_1'}{\H^3}\Big(\frac{a''}{a}-\H\Big)
+\frac{\psi_1^3}{\H^3}\Big[\rhob'''-3\rhob''\Big(\frac{a''}{a}+3\H\Big)+3\rhob'\H\Big(3\frac{a''}{a}-\H\Big)
+\rhob'\Big(\left(\frac{a''}{a}\right)^2-\frac{a'''}{a}\Big)\Big]
 \,.
\end{align}

\subsubsection{Curvature perturbation on uniform density hypersurfaces}
\label{sec:zeta}

The curvature perturbation on uniform density hypersurfaces, $\zeta$, is defined as
\footnote{The sign is taken so as to coincide with $\zeta$ as defined in Ref.~\cite{Bardeen:1983qw}.}
\be
-\zeta\equiv\widetilde{\psi_{\udg}}\,.
\ee
This is obtained by substituting the temporal gauge transformation
components in the uniform curvature gauge (given in section
\ref{sec:alphas}) into the appropriate transformation equation (that
is, Eqs.~(\ref{transpsi1}), (\ref{transpsi2ls}) or
(\ref{transpsi3ls})).  Evaluating this on spatially flat hypersurfaces
then gives, to first, second and third order, respectively,
\be
\zeta_1=-\H\frac{\drho}{\rhob'}\,,
\ee
\be
\zeta_2 = -\frac{\H}{\rhob'}\drhorho+2\H\frac{\drho\drho'}{\rhob'^2}
-\Big[\H\frac{\rhob''}{\rhob'}-\Big(\frac{a''}{a}+\H^2\Big)\Big]\Big(\frac{\drho}{\rhob'}\Big)^2\,,
\ee
\begin{align}
\label{eq:z3cm}
\zeta_3 &= -\H\frac{\drhorhorho}{\rhob'}+\frac{3\H}{\rhob'^2}(\drhorho'\drho+\drho'\drhorho)
-\frac{3}{\rhob'^2}\drhorho\drho\Big[\frac{\H\rhob''}{\rhob'}-\Big(\frac{a''}{a}+\H^2\Big)\Big]
-\frac{3\H}{\rhob'^3}\drho^2\drho''-\frac{6\H}{\rhob'^3}\drho'^2\drho\nn\\
&-\frac{3}{\rhob'^3}\drho^2\drho'\Big[2\Big(\frac{a''}{a}+\H^2\Big)-3\H\frac{\rhob''}{\rhob'}\Big]
-\frac{\drho^3}{\rhob'^3}\Big[3\H\left(\frac{\rhob''}{\rhob'}\right)^2-\H\frac{\rhob'''}{\rhob'}
+\frac{a'''}{a}+3\H \frac{a''}{a}-3\frac{\rhob''}{\rhob'}\Big(\frac{a''}{a}+\H^2\Big)\Big]\,.
\end{align}

There are different definitions of the curvature perturbation present
in the literature, depending on different decompositions of the
spatial part of the metric tensor. A different definition to the
one above, as discussed in e.g.~Ref.~\cite{MW2008}, was used by
Maldacena in Ref.~\cite{Maldacena} to calculate the non-gaussianity
from single field inflation, and was introduced by Salopek and Bond in
Ref.~\cite{Salopek:1990jq}. 
They define the local scale factor
$\tilde{a}\equiv e^{\alpha}$, then
\be
e^{2\alpha}=a^2(\eta)e^{2\zeta}=a^2(\eta)(1+2\zeta_{\rm{SB}}+2\zeta_{\rm{SB}}^2+\frac{4}{3}\zeta_{\rm{SB}}^3)\,.
\ee
Comparing to the expansion from perturbation theory
\be
e^{2\alpha}=a^2(\eta)(1+2\zeta)\,,
\ee
one can obtain the relationship
\be
\zeta=\zeta_{\rm{SB}}+\zeta_{\rm{SB}}^2+\frac{2}{3}\zeta_{\rm{SB}}^3\,.
\ee
Splitting this up order by order gives, at second order
\be
\label{zeta2SB}
\zeta_{\rm{2SB}}=\zeta_2-2(\zeta_1)^2\,,
\ee
and, at third order,
\be
\label{zeta3sb}
\zeta_{3\rm{SB}}=\zeta_{3}-6\zeta_{2}\zeta_1+8(\zeta_1)^3\,.
\ee
Note that it is this definition, (\ref{zeta3sb}), of the curvature
perturbation which occurs in Ref.~\cite{Lehners:2009ja}, though with
different pre-factors since their perturbative expansion is defined
differently\footnote{It is worth mentioning that $\zeta_{\rm{SB}}$,
  the variable first introduced by Salopek and Bond and then employed
  for non-gaussianity calculations by Maldacena, is extremely Gaussian
  after slow-roll inflation, as opposed to other $\zeta$ variables
  which exhibit non-gaussianity, as can be seen e.g.~from
  \eq{zeta2SB}. We thank the anonymous referee for highlighting this
  issue.}.

\section{Governing equations}
\label{sec:equ}

Having constructed gauge invariant quantities up to third order in the
previous section, we now turn to the evolution and the field
equations.

\subsection{Fluid conservation equation}

In this subsection, we give the energy momentum conservation equations
for a fluid with non-zero pressure and in the presence of scalar,
vector and tensor perturbations. The latter generalisation is
important since at orders above linear order, all types of
perturbation are coupled. It is also important to consider rotational
fluids, since second order vorticity in the early universe is likely
non-zero (see, for example, Ref.~\cite{vorticity}). This agrees with
Ref.~\cite{Hwang:2007wq}, but is more general.  \\

The Bianchi identities imply, through the Einstein field equations
which relate the geometry of the space-time to its matter content,
energy momentum conservation,
\be
\label{eq:EMconservation}
\nabla_{\mu}T^{\mu}{}_{\nu}=0,
\ee
where $\nabla_\mu$ here denotes the covariant derivative. Substituting
the definition for the energy momentum tensor (\ref{eq:EMfluid}) into
Eq.~(\ref{eq:EMconservation}) gives energy conservation (the zero
equation)
\begin{align}
\label{eq:energycons}
&\delrho' +3\H(\delrho+\delP)+(\rhob+\Pb)(C^i{}_{i}'+v_{i,}{}^i)+(\delrho+\delP)(C^i{}_{i}'+v_{i,}{}^i)
+(\delrho+\delP)_{,i}v^i \nn \\
&+(\rhob+\Pb)\Big[(B^i+2v^i)(v_i'+B_i')+v^i{}_{,i}\phi-2C_{ij}'C^{ij}
+v^i(C^j{}_{j,i}+2\phi_{,i})+4\H v^i(B_i+2v_i)\Big] \nn \\
&+ (\delrho+\delP)\Big[(B^i+2v^i)(v_i'+B_i') - 2C_{ij}'C^{ij}+v^i(C^j{}_{j,i}
+2\phi_{,i}+4\H(v_i+B_i))+v^i{}_{,i}\phi\Big]  \nn\\
& +(\delrho'+\delP')v^i(B_i+v_i)+(\delrho+\delP)_{,i}\phi v^i +(\rhob+\Pb)(2C^{ij}v_iv_j-B_iv^i\phi+v^iv_i\phi) \nn\\
&
-(\rhob+\Pb)\Big[2C^{ij}(C_{ij,k}v^k-2v_i'v_j +B_iB_j'-2C_i{}^kC_{jk}') 
+\frac{1}{2}v^i{}_{,i}(\phi^2-v^jv_j)\nn \\
& +(v^i+B^i)(2B_i'\phi-C^j{}_j'v_i) 
 -v^j\Big\{B^iB_{i,j}+v^iv_{i,j}+\phi(v_j+2v_j'+3\H v_j-2\phi_{,j}+C^i{}_{i,j}) \nn \\
 &-2\phi'B_j+3C_{ij}'v^i\Big\}
+B^j(B^iC_{ij}'+B_j\phi'+v_j'\phi)\Big] =0\,,
\end{align}
and momentum conservation (the $i-$component)
\begin{align}
&\Big[(\rhob+\Pb)(v_i+B_i)\Big]'+(\rhob+\Pb)(\phi_{,i}+4\H(v_i+B_i))
+\delP_{,i}+\Big[(\delrho+\delP)(v_i+B_i)\Big]'\nn \\
&+(\delrho+\delP)(\phi_{,i}+4\H(v_i+B_i)) -(\rhob'+\Pb')\Big[(2B_i+v_i)\phi -2C_{ij}v^j\Big]\nn\\
&
+(\rhob+\Pb)\Big[(v_i+B_i)(C^j{}_{j}'+v^j{}_{,j})-B_i(\phi'+8\H\phi)
+v^j(B_{i,j}-B_{j,i}+v_{i,j}+8\H C_{ij}) \nn \\
& +(2C_{ij}v^j)'-\phi(v_i'+2B_i'+2\phi_{,i}+4\H v_i)\Big]
+(\rhob'+\Pb')\Big[v^j(B_{i}v_j+B_jv_i+2B_iB_j+\frac{1}{2}v_iv_j\nn\\
&-2C_{ij}\phi)
+(\frac{3}{2}v_i+4B_i)\phi^2\Big]
+(\rhob+\Pb)\Big[\phi^2\left(4B_i'+\frac{3}{2}v_i'+2\H(8B_i+3v_i)+4\phi_{,i}\right) \nn \\
& 
+(v_i+2B_i)(v_j'B^j-C^j{}_j'\phi)+(v_i+B_i)(B_j'B^j-C_{jk}'C^{jk})
+v^j\Big\{2B_i(B_j'+v_j') \nn \\
& +B_j(2B_i'+2\phi_{,i}+v_i')+2\H(v_i+2B_i)(2B_j+v_j)
+2C_{ij}(C^k{}_{k}'+v^k{}_{,k}-4\H\phi)\nn \\
&+C^k{}_{k,j}(B_i+v_i) +v_i(B_j'+\phi_{,j})+v_j(B_i'+\phi_{,i})+v^k(2C_{ik,j}-C_{jk,i})\nn\\
&
+(B_{j,i}-B_{i,j}-2C_{ij}')\phi+2C_{ik}v^k{}_{,j}\Big\}
+\frac{1}{2}(v_iv_jv^j)'-2C_{ij}v^j{}'\phi+B_i(4\phi'\phi-v_{j,}{}^j\phi)\Big] =0\,.
\end{align}

Equation~(\ref{eq:energycons}) highlights the coupling between tensor and
scalar perturbations which occurs only at third order (and higher) in
perturbation theory. At both linear and second order, no such coupling
exists, since the only terms coupling the spatial metric perturbation,
$C_{ij}$, to scalar perturbations contain either the trace or the
divergence of $C_{ij}$ and the tensor perturbation, $h_{ij}$ is, by
definition, transverse and trace-free. However, at third order, terms
like $\delrho C_{ij}'C^{ij}$ occur in the energy conservation equation
which, on splitting up order by order and decomposing $C_{ij}$ becomes
$\drho h_{1ij}'h_1{}^{ij}$. It is clear that this term only shows up
at third order and beyond. Thus, as mentioned earlier, third order is
the lowest order at which all the different types of perturbations
couple to one another in the evolution equations, which will produce
another physical signature of the full theory.

\subsubsection{Scalars only}

It will be useful to have energy and momentum conservation equations
for only scalar perturbations. These equations are obtained by making
the appropriate substitutions $C_{ij}=-\dij\psi+E_{,ij}$,
$v_i=v_{,i},$ and $B_i=B_{,i}$ into the above expressions.  On doing
so, we obtain the energy conservation equation
\begin{align}
&\delrho' +3\H(\delrho+\delP)+(\rhob+\Pb)(\nabla^2v+\nabla^2E-3\psi')+(\delrho+\delP)(\nabla^2v+\nabla^2E-3\psi')
 \nn \\
&+(\delrho+\delP)_{,i}v_,{}^i+(\rhob+\Pb)\Big[(B^i+2v_,{}^i)(v_{,i}'+B_{,i}')+\nabla^2v\phi-2\Big(\psi'(3\psi-\nabla^2E)\nn\\
&
-\psi\nabla^2E'+E_{,ij}'E_,{}^{ij}\Big)+v_,{}^i(2\phi_{,i}+\nabla^2E_{,i}-3\psi_{,i})+4\H v_,{}^i(B_{,i}+2v_{,i})\Big] \nn \\
&+ (\delrho+\delP)\Big[(B_,{}^i+2v_,{}^i)(v_{,i}'+B_{,i}') -2\Big(\psi'(3\psi-\nabla^2E)-\psi\nabla^2E'+E_{,ij}'E_,{}^{ij}\Big)\nn\\
&+v_,{}^i(2\phi_{,i}-3\psi_{,i}+\nabla^2E
+4\H(v_{,i}+B_{,i}))+\nabla^2v\phi\Big] +(\delrho'+\delP')v_,{}^i(B_{,i}+v_{,i})\nn\\
& +(\delrho+\delP)_{,i}\phi v_,{}^i  +(\rhob+\Pb)(v_,{}^iv_{,i}\phi-B_{,i}v_,{}^i\phi-6\psi v_,{}^jv_{,j}+E_,{}^{ij}v_{,i}v_{,j}) \nn\\
&
-(\rhob+\Pb)\Big[2E_,{}^{ij}(v_,{}^kE_{,ijk}-2v_{,i}'v_{,j}+B_{,i}B_{,j}'-2E_{,i}{}^kE_{,kj})-2\nabla^2E\psi_{,k}v_,{}^k\nn\\
&
-2\psi\Big(v_,{}^k(\nabla^2E_{,k}-3\psi_{,k})-2v_{,j}'v_,{}^j+B_{,j}'B_,{}^j+6\psi\psi'\Big)+\frac{1}{2}\nabla^2v(\phi^2-v_,{}^jv_{,j})\nn\\
&
 +(v_,{}^i+\nabla^2E'+B_,{}^i)(2B_{,i}'\phi+3\psi'v_{,i})  -v_,{}^j\Big\{B_,{}^iB_{,ij}+v_,{}^iv_{,ij}
 -2\phi'B_{,j}-3\psi'v_{,j}\nn \\
& 
+\phi(v_{,j}+2v_{,j}'+3\H v_{,j}-2\phi_{,j}-3\psi_{,j}+\nabla^2E_{,j}) +E_{,ij}'v_,{}^i\Big\}\nn\\
&
+B_,{}^j(B_{,j}\psi'+E_{,ij}'v_,{}^i+B_{,j}\phi'+v_{,j}'\phi)\Big] =0\,,
\end{align}

\ni and the momentum conservation equation

\begin{align}
& \Big[(\rhob+\Pb)(v_{,i}+B_{,i})\Big]'+(\rhob+\Pb)(\phi_{,i}+4\H(v_{,i}+B_{,i}))
+\delP_{,i}+\Big[(\delrho+\delP)(v_{,i}+B_{,i})\Big]'\nn \\
& +(\delrho+\delP)(\phi_{,i}+4\H(v_{,i}+B_{,i}))-(\rhob'+\Pb')\Big[(2B_{,i}+v_{,i})\phi+2\psi v_{,i}-2E_{,ij}v_,{}^j\Big]\nn\\
&
+(\rhob+\Pb)\Big[(v_{,i}+B_{,i})(\nabla^2v-3\psi'+\nabla^2E')-B_{,i}(\phi'+8\H\phi)
+v_,{}^j\Big(v_{,ij}-8\H (\psi\dij-E_{,ij}) \Big)
\nn\\
 & -2(\psi v_{,i}+E_{,ij}v_,{}^j)'-\phi(v_{,i}'+2B_{,i}'+2\phi_{,i}+4\H v_{,i})\Big]
 +(\rhob'+\Pb')\Big[v_,{}^j(B_{,i}v_{,j}+B_{,j}v_{,i}+2B_{,i}B_{,j}\nn\\
 &+\frac{1}{2}v_{,i}v_{,j}+2\phi(\psi\dij-E_{,ij}))
+(\frac{3}{2}v_{,i}+4B_{,i})\phi^2\Big] 
 +(\rhob+\Pb)\Big[(v_{,i}+2B_{,i})\Big(v_{,j}'B_,{}^j-(3\psi'-\nabla^2E')\phi\Big)\nn\\
 &
+\phi^2\left(4B_{,i}'+\frac{3}{2}v_{,i}'+2\H(8B_{,i}+3v_{,i})+4\phi_{,i}\right)+(v_{,i}+B_{,i})\Big(B_{,j}'B_,{}^j-3\psi'\psi+\psi'\nabla^2E
\nn \\
& +\psi\nabla^2E'-E_,{}^{jk}E_{,jk}'\Big) +v_,{}^j\Big\{2B_{,i}(B_{,j}'+v_{,j}')+B_{,j}(2B_{,i}'+2\phi_{,i}+v_{,i}')+2\H(v_{,i}+2B_{,i})(2B_{,j}+v_{,j})
\nn \\
& -(3\psi_{,j}-\nabla^2E_{,j})(B_{,i}+v_{,i})-2(\psi\dij-E_{,ij})(\nabla^2v+\nabla^2E'-3\psi'-4\H\phi)+v_{,i}(B_{,j}'+\phi_{,j})\nn\\
&+v_{,j}(B_{,i}'+\phi_{,i})-2\psi_{,j}v_{,i}+E_{,ikj}v_,{}^k+\psi_{,i}v_{,j}
+2(\psi'\dij-E_{,ij}))\phi-2\psi v_{,ij}+2E_{,ik}v_,{}^k{}_j\Big\}\nn\\
&
+\frac{1}{2}(v_{,i}v_{,j}v_,{}^j)'
+2\psi v_{,i}'\phi-2E_{,ij}v_,{}^j{}'\phi+B_i(4\phi'\phi-\nabla^2v\phi)\Big] =0\,.
\end{align}
Note that, in the above, $\nabla^2$ denotes the Laplacian; i.e. $\nabla^2=\partial_{j}\partial^{j}$.

Considering the large scale limit, in which spatial gradients vanish, the energy conservation equation 
becomes
\begin{align}
&\delrho' +3\H(\delrho+\delP)-3\psi'(\rhob+\Pb)-3\psi'(\delrho+\delP)\nn\\
&-6\psi\psi' (\rhob+\Pb) -6\psi\psi'\ (\delrho+\delP) +12\psi^2\psi'(\rhob+\Pb)=0\,.
\end{align}
Splitting up perturbations order by order, this becomes
\begin{align}
&\drhorhorho'+3\H(\drhorhorho+\dPPP)-3\psi_3'(\rhob+\Pb)-9\psi_2'(\drho+\dP)
-9\psi_1'(\drhorho+\dPP)\nn\\
&\quad-18(\rhob+\Pb)(\psi_2\psi_1'+\psi_1\psi_2')+72\psi_1^2\psi_1'(\rhob+\Pb)=0\,.
\end{align}
In the uniform curvature gauge, where $\psi=0$, this is
\begin{align}
&\delta\rho_{3\fg}'+3\H(\delta\rho_{3\fg}+\delta P_{3\fg})=0\,,
\end{align}
and in the uniform density gauge, where $\delta\rho=0$,
\begin{align}
\label{eq:udgzetaev}
3\H\delta P_{3\udg}+3\zeta_3'(\rhob+\Pb)+9\zeta_2'\delta P_{1\udg}
+9\zeta_1'\delta P_{2\udg}
-18(\rhob+\Pb)(\zeta_2\zeta_1'+\zeta_1\zeta_2')
-72\zeta_1^2\zeta_1'(\rhob+\Pb)=0\,,
\end{align}
with $\zeta$ as defined in section \ref{sec:zeta}. This can be recast in the more familiar form by introducing the (gauge invariant) non-adiabatic pressure perturbation. At linear order the pressure perturbation can be expanded as 
\begin{align}
\delta P_1 &=\frac{\p P}{\p S}\delta S_1+\frac{\p P}{\p \rho}\delta\rho_1\equiv\delta P_{\rm{nad}1}+\cs\drho\,.
\end{align}
This can be extended to second order \cite{nonad} and higher by simply not truncating the Taylor series:
\begin{align}
\dPn_2 &= \dPP-\cs\drhorho-\frac{\p \cs}{\p \rho}\drho^2\,,\\
\dPn_3&= \dPPP-\cs\drhorhorho-3\frac{\p \cs}{\p \rho}\drhorho\drho-\frac{\p^2\cs}{\p\rho^2}\drho^3\,.
\end{align}
Thus, in the uniform density gauge, the pressure perturbation is equal
to the non-adiabatic pressure perturbation at all orders. Then,
Eq.~(\ref{eq:udgzetaev}) becomes
\be
\label{zeta3_cons}
\zeta_3'+\frac{\H}{\rhob+\Pb}\dPn_3=
6(\zeta_2\zeta_1'+\zeta_1\zeta_2')+24\zeta_1^2\zeta_1'-\frac{3}{\rhob+\Pb}(\zeta_2'\dPn_1+\zeta_1'\dPn_2)\,.
\ee
In the case of a vanishing non-adiabatic pressure perturbation,
$\zeta_1'$ and $\zeta_2'$ are zero and hence we see that $\zeta_3$ is
also conserved, on large scales. This was also found in
Ref.~\cite{Lehners:2009ja}, and previously in Ref.~\cite{Enqvist:2006fs}.
%

\subsection{Klein Gordon equation}

The energy momentum tensor for a canonical scalar field minimally
coupled to gravity is easily obtained by treating the scalar field as a
perfect fluid with energy-momentum tensor \cite{LLBook}
\be
T^{\mu}{}_{\nu} = g^{\mu\lambda}\vp_{,\lambda}\vp_{,\nu}
-\delta^{\mu}{}_{\nu}\left(\frac{1}{2}g^{\alpha\beta}\vp_{,\alpha}\vp_{,\beta}+U(\vp)\right) \,,
\ee
where the scalar field $\vp$ is split to third order as 
\be
\vp(\eta,x^i)=\vpb(\eta)+\dvp[1](\eta,x^i)+\frac{1}{2}\dvp[2](\eta,x^i)
+\frac{1}{3!}\dvp[3](\eta,x^i) \,,
\ee
and the potential $U$ similarly as 
\be
U(\vp)=U_{0}+\dU[1]+\frac{1}{2}\dU[2]+\frac{1}{3!}\dU[3]\,,
\ee
where we define
\begin{align}
\dU[1] = U_{,\vp}\dvp[1] \,, \qquad
\dU[2] =  U_{,\vp\vp}\dvp[1]^2 + U_{,\vp}\dvp[2] \,, \qquad
\dU[3] = U_{,\vp\vp\vp}\dvp[1]^3+2U_{,\vp\vp}\dvp[1]\dvp[2]
+U_{,\vp}\dvp[3]\,,
\end{align}
and making use of  the shorthand notation $U_{,\vp}\equiv\frac{\p U}{\p\vp}$. Then, Eq. (\ref{eq:EMconservation})
gives the Klein Gordon equation
\begin{align}
\label{eq:KG}
&\dvp[3]''-\nabla^2\dvp[3]+4\H\dvp[3]'+\frac{\vpb''}{\vpb'}\dvp[3]'
-\frac{3\dvp[2]''}{\vpb'}\left(2\vpb'\phi-\dvp[1]'\right)
-\frac{3}{\vpb'}\left(\nabla^2\dvp[2]\dvp[1]'+\dvp[2]'\nabla^2\dvp[1]\right)  \nn\\
&-\frac{6\dvp[2]'}{\vpb'}\left(\phi\vpb''-2\H\dvp[1]'+\vpb\phi'+\vpb'B^{i}{}_{,i}-C^i{}_{i}'\vpb'+4\H\vpb'\phi\right)
-\frac{6(\dvp[1]')^2}{\vpb'}\left(\phi'+B^{i}{}_{,i}+4\H\phi-C^i{}_{i}'\right)\nn \\
&
-\frac{6\dvp[1]''}{\vpb'}\left(2\phi\dvp[1]'-4\vpb'\phi^2+2\vpb'\phi-\frac{1}{2}\dvp[2]'+B^iB_i\vpb'\right)
-\frac{6\dvp[1]'}{\vpb'}\Big[\vpb''(2\phi-4\phi^2+B^iB_i) \nn\\
&+8\phi\vpb'(\H-2\H\phi-2\phi') +B^i(\vpb'B_i'+2\delta\vp_{1,i}'+4\H B_i)
+2\vpb'\Big\{B^i{}_{,i}+\phi'-C^i{}_{i}'(1-2\phi)\nn\\
&-2C^{ij}(B_{j,i}-C_{ij}')+B^iC^j{}_{i,j}-2C^{ij}{}_{,j}B_i
-B^i\phi_{,i}-2B^i{}_{,i}\phi\Big\}-2C^{ij}\delta\vp_{1,ij}\nn\\
&+\delta\vp_{1,i}(B^i{}'+2\H B^i+2C^{ij}{}_{,j}+\phi_,{}^i)\Big]-3\delta\vp_{2,i}\left(B^i{}'+C^j{}_{j,}{}^i+\phi_,{}^i-2C^{ij}{}_{,j}+2\H B^i\right)\nn \\
&-6\delta\vp_{1,i}\Big[B^i{}'+C^j{}_{j,}{}^i-2C^{ij}{}_{,j}+\phi_{,}{}^i+\frac{C^j{}_{j,}{}^i}{\vpb'}
+2\H B^i-2B_jC^{ij}{}'+B^iC^j{}_{j}'+B^jB_{j,}{}^i-B^jB^i{}_{,j} \nn\\
&-B^iB^j{}_{,j}-2C^{ij}B_{j}'-2C^{ij}C^k{}_{k,j}+4C^{ij}C_{jk,}{}^k-2C^{ij}\phi_{,j}-2B^i{}'\phi-B^i\phi'-2\phi_,{}^i\phi
-4\H B_jC^{ij} \nn\\
&-4\H B^i\phi+4C^{kj}C_j{}^i{}_{,k}-2C^{kj}C_k{}^i{}_{,j}\Big]
-12B^i\left(\delta\vp_{1,i}'+\frac{1}{2}\delta\vp_{2,i}'\right)
+24\delta\vp_{1,i}'(B^i\phi+C^{ij}B_j)\nn\\
&
+6\delta\vp_{2,ij}C^{ij}
-6\delta\vp_{1,ij}(4C^{kj}C_{k}{}^{i}-B^iB^j)
-24\H\vpb'\phi(1-2\phi+4\phi^2-\phi'+3\phi'\phi) \nn\\
&-6\vpb''\Big(2\phi(1-2\phi+4\phi^2)+B^iB_i-4B^iB_i\phi-2B^iB^jC_{ij}\Big)
-6C^{ij}{}_{,j}\vpb'(B_i\phi-2B_i)
 \nn\\
&
+6C^i{}_{i}'\vpb'(1-2\phi+4\phi^2-B^jB_j)
-6C^j{}_{j,i}\vpb'(B^i-2C^{ik}B_k-2B^i\phi)
-6C^{ij}{}'\vpb'(2C_{ij}-B_iB_j\nn \\
&-4C_{ij}\phi-4C_{kj}C^k{}_i)
-12\vpb'C^{ij}\Big[2B_iC_{jk,}{}^k+2B^kC_{ki,j}-B_{j,i}-B_iB_j'+B_i\phi_{,j}
-B_kC_{ji,}{}^k\nn \\
&+2C_{ik}B^k{}_{,j}+2B_{j,i}\phi-4\H B_iB_j\Big]
-6\vpb'\Big[\phi+B^i{}_{,i}+B^iB_i'-B^iB_iB^j{}_{,j}-B^iB_jB^j{}_{,i}-4B^iB_i'\phi \nn\\
&+2\H B^iB_i-2B^iB_i\phi'-8\H B^iB_i\phi+4B^i\phi_{,i}-2B^i{}_{,i}\phi+4B^i{}_{,i}\phi^2-B^i\phi_{,i}\Big]
 +6U_{,\vp}a^2=0\,.
\end{align}
One can again see the coupling between first order tensor and scalar perturbations. For example, 
the $\delta\varphi_{1,i}C^{ij}\phi_{,j}$ contains a term that looks like $\delta\varphi_{1,i}h_1{}^{ij}\phi_{1,j}$, which 
occurs only at third order and beyond.

Again, we refrain from splitting up the perturbations order by order for ease of presentation. Once split up,
one can then replace the metric perturbations by using the appropriate order field equations.
We present the Einstein tensor at third order in the next section. Note also that Eq.~(\ref{eq:KG}) implicitly contains the
Klein Gordon equations at first and second order. We refer the reader to, for example, Ref.~\cite{Malik:2006ir}, for
a detailed exposition of the second order Klein Gordon equation.

\subsection{Einstein tensor}

The Einstein tensor, which describes the geometry of the universe, is defined as
\be
G^\mu{}_\nu = R^\mu{}_\nu-\frac{1}{2}\delta^\mu{}_\nu R\,,
\ee
where $R^\mu{}_\nu$ is the Ricci curvature tensor and $R$ is the Ricci scalar. 
Here, we give the components of the Einstein tensor up to third order:

\begin{align}
a^2G^{0}{}_{0} &= -3\H^2+\nabla^2C^{j}{}_{j}-C_{ij,}{}^{ij}
+2\H( -C^{i'}{}_{i}+B^{i}{}_{,i}+3\H\phi)
+C^{j}{}_{j,i}(\frac{1}{2}C^{k}{}_{k,}{}^{i}-2C^{ik}{}_{,k}) +C_{ij}^{'}(\frac{1}{2}C^{ij'}-B^{j}{}_{,}{}^{i})\nn \\
& +B^{i}\Big[ C^{j'}{}_{j,i}-C_{ij,}^{'}{}^{j}
+\frac{1}{2}\left(\nabla^2B_{i}-B_{j,i}{}^{j}\right)+2\H\left(C^{j}{}_{j,i}-2C_{ij,}{}^j-\phi_{,i}\right)\Big]
 +2C^{ij}\Big[2C_{jk,i}{}^{k}-C^{k}{}_{k,ij}-\nabla^2C_{ij} \nn \\
& 
+2\H(C_{ij}^{'}-B_{i,j})\Big]
+C_{jk,i}(C^{ik}{}_{,}{}^{j}-\frac{3}{2}C^{jk}{}_{,}{}^{i}) +C^{i'}{}_{i}(B_{j,}{}^{j}-\frac{1}{2}C^{j'}{}_{j}+4\H\phi)
+2C^{ij}{}_{,i}C_{jk,}{}^{k}\nn \\
&
+\frac{1}{4}B_{j,i}(B^{i}{}_{,}{}^{j}+B^{j}{}_{,}{}^{i})  -3\H^2(4\phi^2-B_{i}B^{i})-\frac{1}{2}B^{i}{}_{,i}B_{j,}{}^{j}
-4\H B^{i}{}_{,i}\phi + \mathbb{G}^0{}_0 \,,
\end{align}

\begin{align}
a^2G^{0}{}_{i} &=  C^{k}{}_{k,i}^{'}
-C_{ik,}^{'}{}^{k} -\frac{1}{2}\left(B_{k,i}{}^k-\nabla^2 B_i\right)-2\H \phi_{,i}
+8\H\phi_{,i}\phi +C_{ij}^{'}\left(2C^{kj}{}_{,k}-C^{k}{}_{k,}{}^{j}+\phi_{,}{}^{j}\right)
-C^{j'}{}_{j}\phi_{,i} \nn \\
& +2C^{kj}\left[C_{ik,j}^{'}-C_{jk,i}^{'}+\frac{1}{2}\left(B_{k,ij}-B_{i,kj}\right)\right]
+B^{j}\left(C_{kj,i}{}^{k}-C^{k}{}_{k,ij}+C_{ik,k}{}^{j}-\nabla^2C_{ij}-2\H B_{j,i}\right)\nn \\
&-\frac{1}{2}\Big(B_{i,j}+B_{j,i}\Big)\phi_{,}{}^{j}  +\left(B_{i,j}-B_{j,i}\right)\left(\frac{1}{2}C^{k}{}_{k,}{}^{j}-C^{jk}{}_{,k}\right)
-C_{ik,j}\left(B^{k}{}_{,}{}^{j}-B^{j}{}_{,}{}^{k}\right)
+B^{j}{}_{,j}\phi_{,i} \nn\\
& +\phi\left[B_{j,i}{}^{j}-\nabla^2B_{i}+2\left(C_{ij,}^{'}{}^{j}-C^{j'}{}_{j,i}\right)\right] -C^{kj'}C_{kj,i}
+\mathbb{G}^0{}_i \,,
\end{align}

\begin{align}
a^2G^{i}{}_{j} &= C^{i''}{}_{j}+2\H C^{i}{}_{j}^{'}-\frac{1}{2}(B^{i'}{}_{,j}+B_{j,}{}^{i'})
-C^{l}{}_{l,j}{}^{i}+C^{i}{}_{l,j}{}^{l}-\phi_{,}{}^{i}{}_{j}-\nabla^2C^{i}{}_{j} 
+C_{jl,}{}^{il}-\H\left(B^{i}{}_{,j}+B_{j,}{}^{i}\right)\nn \\
& +\delta^{i}{}_{j}\left\{\left(\H^2-\frac{2a''}{a}\right)\left(1-2\phi\right)
+2\H\left(B^{k}{}_{,k}-C^{k'}{}_{k}+\phi^{'}\right) +B^{k'}{}_{k}-C^{kl}{}_{,kl}-C^{k''}{}_{k}
+\nabla^2\left(\phi+C^{l}{}_{l}\right)\right\} \nn\\
& +B^{k}\Big[C_{jk,}{}^{i'}+C^{i'}{}_{k,j}-2C^{i'}{}_{j,k}+2\H(C_{jk,}{}^{i}+C^{i}{}_{k,j}-C^{i}{}_{j,k})
+\frac{1}{2}\left(B_{j,}{}^{i}{}_{k}+B^{i}{}_{,jk}-2B_{k,}{}^{i}{}_{j}\right)\Big]\nn\\
&
+(C^{k}{}_{k}^{'}-\phi^{'}-B^{k}{}_{,k})(C^{i}{}_{j}^{'}
-\frac{1}{2}\left(B^{i}{}_{,j}+B_{j,}{}^{i}\right))+C^{ik'}\left(B_{j,k}-2C_{kj}^{'}\right)+C_{kj}^{'}B^{i}{}_{,}{}^{k} +\phi_{,}{}^{i}\phi_{,j}   \nn \\
& +(B^{k'}-2C^{kl}{}_{,l}+C^{l}{}_{l,}{}^{k}+\phi_{,}{}^{k})(C_{jk,}{}^{i}+C^{i}{}_{k,j}-C^{i}{}_{j,k})
+ \frac{1}{2}B^{i}(B_{k,j}{}^{k}-\nabla^2B_{j}+4\H\phi_{,j}-2C^{k'}{}_{k,j}+2C^{'}_{kj,}{}^{k})
\nn \\
& +2C^{ik}\Big[\frac{1}{2}\left(B_{j,k}^{'}+B_{k,j}^{'}\right)-C_{kj}^{''}+\phi_{,jk} -C_{kl,j}{}^{l}-C_{jl,k}{}^{l}
+\nabla^2C_{kj}+C^{l}{}_{l,jk}+\H\left(B_{j,k}+B_{k,j}-2C_{kj}^{'}\right)\Big] \nn \\
& 
-\frac{1}{2}\left(B_{k,}{}^{i}B^{k}{}_{,j}+B_{j,}{}^{k}B^{i}{}_{,k}\right) +\phi\Big[(B_{j,}{}^{i'}+B^{i}{}_{,j}^{'}+2\phi_{,}{}^{i}{}_{j}+2\H(B_{j,}{}^{i}+B^{i}{}_{,j})-2C^{i}{}_{j}^{''}
 -4\H C^{i}{}_{j}^{'}\Big]\nn \\
& +2\left(C^{i}{}_{k,l}C^{k}{}_{j,}{}^{l}-C^{l}{}_{j,}{}^{k}C^{i}{}_{k,l}+C^{kl}{}_{,j}C_{kl,}{}^{i}\right)
+2C^{kl}\Big[C_{kl,j}{}^{i}-C_{jl,}{}^{i}{}_{k}-C^{i}{}_{l,jk}+C^{i}{}_{j,kl}\Big] +\mathbb{G}_o^i{}_j\nn \\
& +\delta^{i}{}_{j}\left\{\left(\H^2-\frac{2a''}{a}\right)(4\phi^2-B_{k}B^{k})
 +2\phi\Big[C^{k''}{}_{k}-B^{k'}{}_{k}-\nabla^2\phi+2\H(C^{k'}{}_{k}-2\phi^{'}-B^{k}{}_{,k})\Big] \right. \nn \\
&\quad +B^{k}\Big[2C^{l'}{}_{l,k}-2C_{kl,}^{'}{}^{l}+\nabla^2B_{k}-B_{l,k}{}^{l}+2\H(B_{k}^{'}-\phi_{,k}-2C^{l}{}_{k,l}
+C^{l}{}_{l,k})\Big]+C^{kl'}\left(\frac{3}{2}C_{kl}^{'}-B_{l,k}\right) \nn \\
& \quad +2C^{kl}\Big[C_{kl}^{''}-\nabla^2C_{kl}+2\H C_{kl}^{'}+2C_{lm,k}{}^{m}-C^{m}{}_{m,kl}
-2\H B_{l,k}-B_{l,k}^{'}-\phi_{,kl}\Big] +2B^{k'}(C^{l}{}_{l,k}-C_{kl,}{}^{l})\nn \\
& \quad +C^{k'}{}_{k}\left(B^{l}{}_{,l}-\frac{1}{2}C^{l'}{}_{l}\right)
+2C^{kl}{}_{,k}C_{lm,}{}^{m} +C_{lm,k}\left(C^{km}{}_{,}{}^{l}-\frac{3}{2}C^{lm}{}_{,}{}^{k}\right)
-C^{l}{}_{l,k}\left(2C^{k}{}_{m,}{}^{m}-\frac{1}{2}C^{m}{}_{m,}{}^{k}\right)\nn \\
& \quad 
+\phi^{'}\left(C^{k'}{}_{k}-B^{k}{}_{,k}\right) \left. -\frac{1}{4}\left(2B^{k}{}_{,k}B_{l,}{}^{l}-B_{l,k}B^{k}{}_{,}{}^{l}
-3B^{l}{}_{,k}B_{l,}{}^{k}\right) + \phi_{,k}\left(C^l{}_{l,}{}^k-2C^{lk}{}_{,l}-\phi_{,}{}^k\right) +\mathbb{G}_d^i{}_j \right\} \,,
\end{align}
where $\mathbb{G}^0{}_0, \mathbb{G}^0{}_i, \mathbb{G}^i{}_j$ are the
third order corrections (the latter split into a diagonal part
$\mathbb{G}^i_o{}_j$, and an off diagonal part $\mathbb{G}^i_d{}_j$)
which we give in the appendix as Eqs. (\ref{eq:00}), (\ref{eq:0i}),
(\ref{eq:ijo}) and (\ref{eq:ijd}), respectively. Note that, in
calculating the third order components given above, we have implicitly
obtained the full second order Einstein tensor components for fully
general perturbations (i.e.~including all scalar, vector and tensor
perturbations). We shall extend this second order analysis, by
presenting all the geometric and matter tensors as well as
conservation and constraint equations in full generality, in a future
publication \cite{inprep}.

\section{Discussion and conclusions}
\label{sec:dis}

In this paper we have developed the essential tools for cosmological
perturbation theory at third order. Starting with the definition of
the active gauge transformation we have extended the work of
Ref.~\cite{MW2008} to third order, and derived gauge invariant
variables, namely the curvature perturbation on uniform density
hypersurfaces, $\zeta_3$, and the density perturbation on uniform
curvature hypersurfaces. We also relate the curvature perturbation
$\zeta_3$, obtained using the spatial metric split of
Ref.~\cite{MW2008} to that introduced by Salopek and Bond
\cite{Salopek:1990jq}, which is also popular at higher order.

We have then calculated the energy and momentum conservation equations
for a general perfect fluid at third order, including all scalar,
vector and tensor perturbations. The Klein Gordon equation for a
canonical scalar field minimally coupled to gravity is also
presented. We highlight the coupling in these conservation equations
between scalar and tensor perturbations which only occurs at third
order and above. 
Finally, we have presented the Einstein tensor components to third
order. No large scale approximation is employed for the tensor
components or the conservation equations. All equations are given
without specifying a particular gauge, and can therefore immediately
be rewritten in whatever choice of gauge is desired. However, as
examples to illustrate possible gauge choices, we give the energy
conservation equation on large scales (and only allowing for scalar
perturbations) in the flat and the uniform density gauge. This gives
an evolution equation for the curvature perturbation $\zeta_3$,
\eq{zeta3_cons}. As might be expected from fully non-linear
calculations \cite{LMS} and second order perturbative calculations
\cite{MW2003}, the curvature perturbation is also conserved at third
order on large scales in the adiabatic case. It is worth noting that 
higher order perturbation theory, as discussed in this paper,
has the advantage of being valid on all scales whereas fully non-linear
methods, such as the $\delta N$ formalism, 
are only valid in the large scale limit.

Another application of our third order variables and equations, in
particular the Klein Gordon equation (\ref{eq:KG}), is the calculation of
the trispectrum by means of the field equations. Whereas calculations
of the trispectrum so far derive the trispectrum from the fourth order
action, it should also be possible to use the third order field
equations instead. The equivalence of the two approaches for
calculating the bispectrum, using the third order action or the second
order field equations, has been shown in Ref.~\cite{Seery:2008qj}. 
Having included tensor as well as scalar perturbations it will be in
particular interesting to see and be an important consistency check
for the theory whether we arrive at the same result as
Ref.~\cite{Seery:2008ax}.

A final advantage of extending perturbation theory to third order is
that, in doing so, one obtains a deeper insight into the second order
theory. Also second order perturbation theory, despite remaining
challenging, becomes less daunting having explored some of the third
order theory.

\acknowledgments

The authors are grateful to Jim Lidsey, David Lyth, Misao Sasaki and
David Wands for useful comments and discussions. AJC is supported by
the Science and Technology Facilities Council (STFC). We used the
computer algebra package {\sc{Cadabra}} \cite{Peeters} to assist with
the calculations, and thank Kasper Peeters for useful discussions on
using the package.

\appendix
\section{Einstein Tensor}
Here, we give the third order corrections to the components of the Einstein tensor. We do not split up perturbations
order by order.

\begin{align}
\label{eq:00}
\mathbb{G}^{0}{}_{0} &= 2C^{ij}\Big[2C_{ik}(C^l{}_{l,j}{}^k-2C^{kl}{}_{,jl}+\nabla^2C_j{}^k-2\H C'_j{}^k)
+(2C_{jk,i}-C_{ij,k})(C^l{}_{l,}{}^k-2C^k{}_{l,}{}^l)+C_{il,k}(3C_j{}^l{}_,{}^k-C_j{}^k{}_{,l}) \nn \\
& +C^k{}_{k,i}(2C^j{}_{l,}{}^l-\frac{1}{2}C^l{}_{l,j})+B^i\Big\{C'_{jk,}{}^k
-C^k{}_{k,j}'+\frac{1}{2}(B_{k,j}{}^k-\nabla^2B_j)+\H(4C_{jk,}{}^k+2\phi_{,j}-3\H B_j  -2C^k{}_{k,j})\Big\}
\nn \\
&
+C_{ik}'(B^k{}_{,j}+B_{j,}{}^k) -2C_{ik,}{}^kC_{jl,}{}^l-(B^k{}_{,k}+\phi)(C_{ij}'-B_{j,i})-\frac{1}{4}B_{k,i}(B^k{}_{,j}+2B_{j,}{}^k)
-\frac{1}{4}B_{i,k}B_{j,}{}^k-C^k{}_{k}'B_{j,i}\nn \\
& +8\H C_{ik}B^k{}_{,j}\Big] + 2C^{kj}B^i\Big[C_{ik,j}'-C_{jk,i}'+\frac{1}{2}(B_{k,ij}-B_{i,jk})+2\H(2C_{ik,j}-C_{kj,i})\Big]
+C^{ij}{}'\Big[2B^k(C_{ki,j}-C_{ji,k}) \nn \\
& +B_i(2C_{jk,}{}^k-C^k{}_{k,j})+B_i\phi_{,j}+(2B_{j,i}-C_{ij}')\phi\Big]+8\H\phi^2(B^i{}_{,i}-C^i{}_{i}')
+(B_{i,j}-B_{j,i})(\frac{1}{2}B^iC^k{}_{k,}{}^j-B^iC^{jk}{}_{,k}) \nn \\
& -\frac{1}{2}(B_{i,j}+B_{j,i})(\phi B^{j}{}_{,}{}^{i}+B^i\phi_{,}{}^j)-2\H B^iB^jB_{j,i}
+B^j{}_{,j}\Big[B^i(2C_{ik,}{}^k-C^k{}_{k,i}+\phi_{,i}-2\H B_i)+B^i{}_{,i}\phi-2C^i{}_{i}'\phi\Big] \nn \\
& +B_{k,j}B^i(C^{kj}{}_{,i}-2C_i{}^k{}_,{}^j) +2\H B^i\Big[C^j{}_{j}'B_i+2\phi(2C_{ij,}{}^j-C^j{}_{j,i}+2\phi_{,i}
-3\H B_i)\Big] +C^i{}_{i}'C^j{}_{j}'\phi\nn \\
& +B^iB^j(2C_{jk,i}{}^k-C^k{}_{k,ij}-\nabla^2C_{ij}) +B^iC^j{}_{j}'(C^k{}_{k,i}-2C_{ik,}{}^k-\phi_{,i})+B^i\phi(B^j{}_{,ij}+2C_{ij,}'{}^j-2C^j{}_{j,i}'-\nabla^2B_i) \,,
\end{align}

\begin{align}
\label{eq:0i}
\mathbb{G}^{0}{}_{i} &= C_{kj}\left[2C^{jl}\left(C^k{}_{l,i}{}'-C_{il,}{}^{k'}\right)
+C_i{}^j\left(C^l{}_{l,}{}^k-2C^{kl}{}_{,l}-2\phi_{,}{}^k\right)+C'_{il}\left(C^{jk}{}_,{}^l-2C^{kl}{}_,{}^j\right)
+2C^{jl}{}'C^k{}_{l,i}\right] \nn \\
& +C^{kj}\Big[(B_{j,i}-B_{i,j})(C^{l}{}_{l,k}-C_{kl,}{}^l)+(B_{l,i}-B_{i,l})(C_{jk,}{}^l-2C_k{}^l{}_{,j})
+2(B_{j,l}-B_{l,j})(C_{ik,}^l-C_i{}^l{}_{,k})+(B_{j,i}+B_{i,j})\phi_{,k}  \nn \\
& + 2(C_{jk}'-B_{k,j})\phi_{,i}+2\phi(2C_{jk,i}'-2C_{ik,j}'+B_{i,kj}-B_{k,ij}) \Big]
+2B_j\Big[C^{kj}(C^l{}_{l,ik}-C_{kl,i}{}^l-C_{il,k}{}^l+\nabla^2C_{ik}) \nn \\
& +C^{kl}(C_{kl,i}{}^j-C^j{}_{l,ik}-C_{il,k}{}^j+C_{i}{}^j{}_{,kl})
+(C^l{}_{l,}{}^k-C^{kl}{}_{,l})(C^j{}_{k,i}+C_{ik,}{}^j-C^j{}_{i,k})
+C_{il,k}(C^{jl}{}_{,}{}^k-C^{jk}{}_{,}{}^l) \nn \\
& +\frac{1}{2}C_{kl,i}C^{kl}{}_{,}{}^{j}\Big]+2C_{ij}'\phi(C^k{}_{k,}{}^j-2C^{kj}{}_{,k})
+B^j\Big[B_{j}(C_{ik,}'{}^k-C^k{}_{k,i}'+\frac{1}{2}(B_{k,i}{}^k-\nabla^2B_i))
+C_{ik}'B_{j,}{}^k \nn \\
& +B_{j,i}(B^k{}_{,k}-C^k{}_{k}') +\frac{1}{2}B_{j,k}(2C_i'{}^k-B^k{}_{,i}B_{i,}{}^k)\Big]
+2\phi\Big[(B_{i,j}-B_{j,i})(C^{kj}{}_{,k}-\frac{1}{2}C^k{}_{k,}{}^j)+\phi_,{}^j(B_{i,j}+B_{j,i}) \nn \\
& +C_{kj}'C^{kj}{}_{,i}+2C^j{}_j'\phi_{,i}-2C_{ij}'\phi_,{}^j-B_{k,j}(C_i{}^j{}_,{}^k-C_i{}^k{}_,{}^j)
-2B_{j,}{}^j\phi_{,i}\Big] +2C_{kj}C^{jl}(B_{i,l}{}^k-B_{l,i}{}^k) \nn \\
& +B^j\Big[\phi_{,i}(C^k{}_{k,j}-2C_{jk,}{}^k+4\H B_j)
+2\phi(C^k{}_{k,ij}-C_{jk,}{}^k{}_i-C_{ik,j}{}^k+\nabla^2C_{ij}+4\H B_{j,i}) +2\H(B^kC_{jk,i}+2C_{kj}B^k{}_{,i}) \nn \\
& +\phi_,{}^k(C_{ij,k}-C_{ik,j}+C_{jk,i})\Big] +2\phi^2(2C^j{}_{j,i}'-2C_{ij,}'{}^j-B_{j,i}{}^j+\nabla^2B_i-12\H\phi_{,i})\,,
\end{align}

\begin{align}
\label{eq:ijo}
\mathbb{G}^{i}_{\rm{o}}{}_{j} &= 
2C^{ik}C_{kl}\Big[2C^l{}_j''-B^l{}_{,j}'-B_{j,}{}^l{}'-2C^m{}_{m,j}{}^l+2C^l{}_{m,j}{}^m
+2C_{jm,}{}^{lm}-2\nabla^2C^l{}_j-2\phi_{,j}{}^l\Big]\nn\\
& +4C^{ik}C^{lm}(C_{km,jl}-C_{lm,kj}+C_{jm,kl}-C_{jk,lm})
+4C^{km}C_{kl}(C_{jm,}{}^{il}-C^l{}_{m,}{}^i{}_j+C^i{}_{m,j}{}^l-C^i{}_{j,}{}^l{}_m)\nn\\
& +C^{ik}\Big[2(2C^{lm}{}_{,m}-C^m{}_{m,}{}^l-\phi_,{}^l-B^l{}')(C_{kl,j}+C_{jl,k}-C_{jk,l})
+4C_{jm,l}(C_k{}^l{}_,{}^m-C_k{}^m{}_,{}^l)-2C_{lm,j}C^{lm}{}_{,k}\nn\\
&-2C_{jk}'(C^l{}_l'-B^l{}_{,l}-\phi')+2C_{jl}'(2C^l{}_k'-B_{k,}{}^l)+C^l{}_l'(B_{k,j}+B_{j,k})
-2C_{kl}'B_{j,}{}^l-(\phi'+B^l{}_{,l})(B_{k,j}+B_{j,k})\nn\\
&+B_{l,j}B^l{}_{,k}+B_{j,l}B_{k,}{}^l-2\phi_{,k}\phi_{,j}-2\phi(B_{k,j}'+B_{j,k}'+\phi_{,jk}
+4\H C_{kl}(2C^l{}_j'-B^l{}_{,j}-B_{j,}{}^l)-4\H B_k\phi_{,j}\Big]\nn\\
&+C^{kl}\Big[2(C^m{}_{m,l}-2C_{lm,}{}^m)(C^i{}_{j,k}-C^i{}_{k,j}-C_{jk,}{}^i)
+2(2C^m{}_{l,k}-C_{kl,}{}^m)(C_{jm,}{}^i+C^i{}_{m,j}-C^i{}_{j,m})-4C_{km,}{}^iC^m{}_{l,j}\nn\\
& +4(C_{jl,m}-C_{jm,l})(C^{im}_{,k}-C^i{}_{k,}{}^m)-2C^i{}_j'C_{kl}'+4C^i{}_k'C_{jl}'
+(B^i{}_{,j}+B_{j,}{}^i)(C_{kl}'-B_{l,k})-2\phi_,{}^l(C_{jk,}{}^i+C^i{}_{k,j}-C^i{}_{j,k})\nn\\
& +2C^i{}_j'B_{l,k}-2C^i{}_k'B_{j,l}-2C_{jk}'B^i{}_{,l}+B_{k,}{}^iB_{l,j}
-2B_k'(C_{jl,}{}^i+C^i{}_{l,j}-C^i{}_{j,l})\Big]
-B^kB_k\Big[C^i{}_j''-\frac{1}{2}(B^i{}_{,j}'+B_{j,}{}^i{}')\nn\\
& -\phi_{,}{}^i{}_j-\H(B^i{}_{,j}+B_{j,}{}^i-2C^i{}_j')\Big]
+B^iB^k{}'\Big[C^l{}_{l,jk}-C^l{}_{k,jl}-C^l{}_{j,kl}+\nabla^2C_{jk}+2\H B_{k,j}\Big]
\nn\\
& +B^kB^l\Big[C_{kl,}{}^i{}_j-C_{jl,}{}^i{}_k-C^i{}_{l,jk}+C^i{}_{j,kl}\Big]
+B^i\Big[C_{jk}'(C^l{}_{l,}{}^k-2C^{kl}{}_{,l}-\phi_,{}^k)+C_{kl}'C^{kl}{}_{,j}
+\frac{1}{2}\phi_{,}{}^k(B_{k,j}+B_{j,k})\nn
\end{align}
\begin{align}
&+\left(\frac{1}{2}C^l{}_{l,}{}^k-C^{kl}{}_{,l}\right)(B_{k,j}-B_{j,k})+B_{l,k}(C^l{}_{j,}{}^k-C^k{}_{j,}{}^l)
-\phi_{,j}(B^k{}_{,k}-C^k{}_{k}')\Big]+4\H\phi^2(2C^i{}_j'-B^i{}_{,j}-B_{j,}{}^i)\nn\\
&+B^k\Big[C^{i}{}_k(2C^l{}_{l,j}'-2C_{jl,}{}^l{}'-B^l{}_{,jl}+\nabla^2B_j)
+C^{il}(4C_{jl,k}'-2C_{kl,j}'-C_{jk,l}'+2B_{k,jl}-B_{l,kj}-B_{j,kl})\nn\\
&+C^l{}_k(4C^i{}_{j,l}'-2C^i{}_{l,j}'-C_{jl,}{}^i{}'+2B_{l,j}{}^i-B_{j,l}{}^i-B^i{}_{,jl})
+C^i{}_j'(2C_{kl,}{}^l-B_k'-C^l{}_{l,k}+\phi_{,k})+2C^{il}{}'(C_{jl,k}-C_{jk,l})\nn\\
&+2C_{jl}'(C^{il}{}_{,k}-C^i{}_{k,}{}^l)+\frac{1}{2}(B_k'+C^l{}_{l,k}-2C_{kl,}{}^l)\left(B_{j,}{}^i+B^i{}_{,j}\right)
+(C^l{}_{j,}{}^i+C^{il}{}_{,j}-C^i{}_{j,}{}^l)(B_{k,l}-B_{l,k}-2C_{kl}'-4\H C_{kl})\nn\\
&+(C_{jk,}{}^i+C^i{}_{k,j}-C^i{}_{j,k})(C^l{}_l'-\phi'-B^l{}_{,l})
+(B^i{}_,{}^l-4\H C^{il})(C_{jk,l}-C_{jl,k})+(B^l{}_,{}^i-4\H C^{il})C_{kl,j}+B^l{}_{,j}C_{kl,}{}^i\nn\\
&-\frac{1}{2}\phi_{,k}(B^i{}_{,j}+B_{j,}{}^i)+B_{j,l}(C^i{}_{k,}{}^l-C^{il}{}_{,k})+B_{k,}{}^i\phi_{,j}+B_{k,j}\phi_,{}^i\Big]
+2\phi\Big[B^i\Big(C^k{}_{k,j}'-C_{jk,}{}^k{}'+\frac{1}{2}(\nabla^2B_j-B^k{}_{,kj})\Big)\nn\\
&+B^k\Big(2C^i{}_{j,k}'-C^i{}_{k,j}'-C_{jk,}{}^i{}'+B_{k,j}{}^i-\frac{1}{2}(B_{j,k}{}^i+B^i{}_{,jk})\Big)
+(B^k{}'+\phi_{,}{}^k+4\H B^k)(C^i{}_{j,k}-C^i{}_{k,j}-C_{jk,}{}^i)\nn\\
&+C^i{}_k'(2C^k{}_j'-B_{j,}{}^k)+2C^{ik}C_{jk}''-C_{jk}'B^i{}_,{}^k
+\frac{1}{2}(B^i{}_{,j}+B_{j,}{}^i)(C^k{}_k'-2\phi'-B^k{}_{,k})+C^i{}_j'(B^k{}_{,k}-C^k{}_k'+2\phi')\nn\\
&+4\phi_{,j}(2\H B^i-\phi_{,}{}^i)-4\H C^{ik}(2C_{kj}'+B_{k,j}+B_{j,k})
+\frac{1}{2}\left(B_{k,}{}^iB^k{}_{,j}+B^i{}_{,k}B_{j,}{}^k\right)\Big]\,, 
\end{align}

\begin{align}
\label{eq:ijd}
\mathbb{G}^{i}_{\rm{d}}{}_{j} &=
4C^{kl}C_{km}\Big[\nabla^2C_l{}^m-C^m{}_l''+C^n{}_{n,l}{}^m-2C^{mn}{}_{,nl}
+B^m{}_{,l}'+\phi_{,l}{}^m+2\H(B^m{}_{,l}-C^m{}_{l}')\Big]\nn\\
&
+C^{kl}\Big[4C_{mn}(C^{mn}{}_{,kl}-2C_l{}^n{}_{,k}{}^m)+2C_{km}'(B^m{}_{,l}+B_{l,}{}^m-3C^m{}_l')
+2(C^n{}_{n,}{}^m-2C^{mn}{}_{,n})(2C_{lm,k}-C_{kl,m})\nn\\
& 
+B^m{}'(4C_{ml,k}-C_{kl,m})+2B_k'(2C_{lm,}{}^m-C^m{}_{m,l})
+C^m{}_{m,k}(4C_{ln,}{}^n-C^n{}_{n,l})+C_{mn,k}(3C^{mn}{}_{,l}-4C^{n}{}_{l,}{}^m)\nn\\
&+C_{kn,m}(6C_l{}^n{}_,{}^m-2C_l{}^m{}_,{}^n)+4C_{lm,k}\phi_,{}^m-2C^m{}_{m,k}\phi_{,l}
-2C_{kl,m}\phi_,{}^m+2\phi_{,k}\phi_{,l}+4C_{km,}{}^m(\phi_{,l}-C_{ln,}{}^n) \nn \\
& +2(B_{l,k}-C_{kl}')(B^m{}_{,m}+\phi'+4\H\phi)+4\phi(\phi_{,kl}+B_{l,k}'-C_{kl}'')
+B_{m,k}(B_{l,}{}^m-\frac{3}{2}B^m{}_{,l})\nn\\
& +2B_{k}\Big(2C_{lm,}{}^k-2C^m{}_{m,l}'+B^m{}_{,lm}-\nabla^2B_l-2\H(B_l'
+C^m{}_{m,l}-2C_{lm,}{}^m-\phi_{,l})\Big)-\frac{3}{2}B_{k,m}B_{l,}{}^m\Big]\nn\\
& +C^k{}_k'\Big[2C^{ml}(C_{ml}'-B_{l,m})+\phi C^l{}_l'-8\H\phi^2-4\phi\phi'-2\phi B^l{}_{,l}
+B^l(B_l'+C^m{}_{m,l}-2C_{lm,}{}^m-\phi_{,l}+2\H B_l)\Big]\nn\\
& +B^k\Big[B_k\Big(C^l{}_l''-B^l{}_{,l}'-\nabla^2\phi-2\H(2\phi'+B^l{}_{,l})\Big)
+B^l\Big(2C_{lm,k}{}^m-C^m{}_{m,kl}-\nabla^2C_{kl}-2\H(C_{kl}'+B_{l,k})\Big) \nn \\
& -\left(\frac{2a''}{a}-\H^2\right)(2C_{kl}+4\phi)
+2C^{lm}\Big(2C_{km,l}'-2C_{lm,k}'+B_{m,lk}-B_{k,ml}-2\H(C_{lm,k}-2C_{km,l}\Big)\Big]\nn\\
& -4\phi^2\Big[C^k{}_k''-B^k{}_{,k}'-\nabla^2\phi+2\H(\H\phi-3\phi'-B^k{}_{,k})\Big]
+C^{kl}{}'\phi\Big(2B_{l,k}-3C_{kl}'\Big)+\frac{1}{2}B_{l,k}\phi(B^k{}_,{}^l-3B^l{}_,{}^k)\nn\\
&+B^k{}_{,k}\phi(4\phi'+B^l{}_{,l})+2\phi(C_{kl,}{}^k\phi_,{}^l-C^l{}_{l,k}\phi_,{}^k+2\phi_{,k}\phi_,{}^k)
+2B^k{}'(2C_{kl,}{}^l\phi-\phi C^l{}_{l,k})\nn\\
&+B^k\Big[C^m{}_{m,}{}^l(B_{k,l}-B_{l,k}-2C_{kl}')+C^{lm}{}_{,m}(4C_{kl}'-2B_{k,l})
+B^l{}_{,l}(2C_{km,}{}^m-C^m{}_{m,k}-B_k'+\phi_{,k})\nn\\
& +B_{m,l}(C^l{}_{k,}{}^m-3C^m{}_{k,}{}^l+C^{lm}{}_{,k})+C_{lm}'(C_k{}^m{}_,{}^l-C^{lm}{}_{,k})
-\phi'(C^l{}_{l,k}-2C_{kl,}{}^l)+B^l{}_{,k}C_{ml,}{}^m \nn \\
& +2\phi\Big(2C_{kl,}{}^l{}'-2C^l{}_{l,k}'+B_{l,k}{}^l-\nabla^2B_k
-4\H(2B_k'+C^l{}_{l,k}-2C_{kl,}{}^l-2\phi_{,k})\Big)\Big].
\end{align}



\end{document}